\documentclass[twocolumn]{aastex631}
\usepackage{CJK}
\usepackage{relsize}
\usepackage{fontawesome}
\usepackage{multirow}
\usepackage{booktabs}
\usepackage{longtable}
\let\tablenum\relax
\shorttitle{Estimating Ion Temperatures at the Polar Coronal Hole Boundary}
\shortauthors{Zhu et al.}
\graphicspath{{./}{figures/}}

\begin{document}

\title{Estimating Ion Temperatures at the Polar Coronal Hole Boundary}

\begin{CJK*}{UTF8}{gbsn}
\author[0000-0003-3908-1330]{Yingjie Zhu (朱英杰)}
\altaffiliation{DKIST Ambassador}
\affiliation{Department of Climate and Space Sciences and Engineering, University of Michigan, \\
Ann Arbor, MI 48109, USA}

\author[0000-0002-9465-7470]{Judit Szente}
\affiliation{Department of Climate and Space Sciences and Engineering, University of Michigan, \\
Ann Arbor, MI 48109, USA}

\author[0000-0002-9325-9884]{Enrico Landi}
\affiliation{Department of Climate and Space Sciences and Engineering, University of Michigan, \\
Ann Arbor, MI 48109, USA}

\correspondingauthor{Yingjie Zhu (朱英杰)}
\email{yjzhu@umich.edu}



\begin{abstract}
The temperatures of the heavy ions ($T_i$) in the solar corona provide critical information about the heating mechanism of the million-degree corona. However, the measurement of $T_i$ is usually challenging due to the nonthermal motion, instrumental limitations, and the optically thin nature of the coronal emissions. We present the measurement of $T_i$ and its dependency on the ion charge-to-mass ratio ($Z/A$) at the polar coronal hole boundary, only assuming that heavy ions have the same nonthermal velocity. To improve the $Z/A$ coverage and study the influence of the instrumental broadening, we used a coordinated observation from the extreme-ultraviolet Imaging Spectrometer (EIS) on board the Hinode satellite and the Solar Ultraviolet Measurements of Emitted Radiation (SUMER) on board the Solar and Heliospheric Observatory (SOHO). We found that the $T_i$ of ions with $Z/A$ less than 0.20 or greater than 0.33 are much higher than the local electron temperature. We ran the Alfv\'en Wave Solar Model-realtime to investigate the formation of optically thin emissions along the line of sight (LOS). The simulation suggested that plasma bulk motions along the LOS broaden the widths of hot emission lines in the coronal hole (e.g., Fe \textsc{xii}, Fe \textsc{xiii}). We discussed other factors that might affect the $T_i$ measurement, including the non-Gaussian wings in some bright SUMER lines, which can be fitted by a double-Gaussian or a $\kappa$ distribution. Our study confirms the preferential heating of heavy ions in coronal holes and provides new constraints on coronal heating models.


\end{abstract}

\keywords{Solar coronal holes(1484), Solar coronal lines (2038), Solar coronal heating (1989), Spectroscopy (1558)}


\section{Introduction} \label{sec:intro}
The heating of the million-degree solar corona above the photosphere has been one of the major mysteries in solar physics research since the 1940s. The energy fluxes needed to heat different coronal structures range from 3$\times 10^5$ (quiet Sun), 8$\times 10^5$ (coronal holes), to $10^7\,\mathrm{erg\,cm^{-2}\,s^{-1}}$ \citep[active regions,][]{Withbroe1977}. Coronal holes are the darkest areas on the solar disk or above the limb in extreme ultraviolet (EUV) or X-Ray images, because of their low density \citep{Cranmer2009}. The open magnetic field structure and the fast solar wind make the coronal hole an excellent laboratory to study the mechanisms of coronal heating and solar wind acceleration, especially the wave dissipation and turbulence models \citep[e.g.,][]{Hollweg2002,Cranmer2007}. Distinguishing the contribution of these proposed mechanisms requires the measurements of electron temperature, ion temperature, and nonthermal motions in the polar coronal hole \citep[e.g.,][]{Wilhelm2012}. 


The thermal width of the spectral line is the only remote-sensing measurement of ion temperatures $T_i$ in the corona \citep{DelZanna2018}. Observations from the Solar Ultraviolet Measurements of Emitted Radiation \citep[SUMER;][]{Wilhelm1995} on board the Solar and Heliospheric Observatory \citep[SOHO;][]{Domingo1995} showed that the spectral lines in the coronal hole below 1.5$\,R_\odot$ are much broader than the profiles in streamers. The ion temperatures of Ne, Mg, Fe, and S are more than 2.5 times higher than their formation temperatures \citep{Seely1997}. In the darkest region of the coronal hole, Si \textsc{viii} and Ne \textsc{viii} show extreme effective temperatures of $10^7$ and $2.3\times 10^7$\,K \citep{Wilhelm1998,Wilhelm1999}. Besides, observations from the Ultraviolet Coronagraph Spectrometer \citep[UVCS;][]{Kohl1995} on board SOHO indicated that O \textsc{vi} and Mg \textsc{x} ions are preferentially heated to $10^7$\textendash$10^8$\,K compared to the protons above the polar coronal hole between 1.35 and $3\,R_\odot$, where ion collisions become infrequent \citep[e.g.,][]{Kohl1997, Esser1999, Doyle1999}. Significant O \textsc{vi} temperature anisotropy perpendicular to the field lines is also found using the Doppler dimming or pumping of O \textsc{vi} 1032/1037\,\mbox{\AA} lines \citep[e.g.,][]{Kohl1998,Li1998}. 

The ion-cyclotron resonance is one of the promising candidates for explaining the preferential and anisotropic heating of heavy ions in the corona \citep[e.g.,][]{Marsch1982,Isenberg1983}. The heating might occur when heavy-ion particles interact with the waves generated through turbulent cascade \citep[e.g.,][]{Hu1999}, or via activity in the chromospheric network \citep[e.g.,][]{Tu1997}, or by the local instability \citep[e.g.,][]{Markovskii2004}. As the wave\textendash particle interaction is sensitive to the gyrofrequency of heavy ions, the ion charge-to-mass ratio ($Z/A$) plays an essential role in determining the heating efficiency of ion-cyclotron resonance \citep[e.g.,][]{Patsourakos2002}.


 However, few studies have focused on the dependence of $T_i$ on $Z/A$ using remote-sensing observations. Most of the studies used observations from SOHO/SUMER and reached different conclusions. \citet{Tu1998,Tu1999} found that $T_i$ remains constant or slightly decreases with increasing $Z/A$ in the polar coronal hole. \citet{Dolla2008,Dolla2009} also reported that $T_i$ decreases with the increase in $Z/A$, but the low $Z/A$ species (i.e., Fe \textsc{viii} and Fe \textsc{x}) are significantly heated. \citet{Wilhelm2005} found a linear relation between $T_i$ and $Z/A$ in the quiet Sun only if the Ca \textsc{xiii} and Fe \textsc{xvii} widths are discarded. \citet{Landi2007} investigated the SUMER quiet-Sun observations during different solar activity levels and concluded no correlation between $T_i$ and $Z/A$. On the other hand, \citet{Landi2009} suggested a nonmonotonic dependence of $T_i$ on $Z/A$ in a coronal hole.

A couple of studies used observations from the EUV Imaging Spectrometer \citep[EIS;][]{Culhane2007} on board the Hinode \citep{Kosugi2007} satellite to study the dependence of $T_i$ on $Z/A$ in different regions. \citet{Hahn2010} found that $T_i$ decreases with $Z/A$ in the off-limb polar coronal hole. However, in the quiet Sun, $T_i$ of different ions appears to be constant \citep{Hahn2014}. \citet{Hahn2013a} study the ion temperature anisotropy in an on-disk coronal hole and found that only the perpendicular ion temperature $T_{i,\perp}$ shows a dependence on $Z/A$. In contrast, the parallel ion temperature $T_{i,\parallel}$ is relatively constant.  

The primary difficulty in measuring $T_i$ is that, apart from the thermal motions, unresolved nonthermal motions also contribute to the observed line width. Therefore, scientists had to make additional assumptions to separate the thermal and nonthermal components in observations. The assumptions include $T_i$ equals the line formation temperature \citep{Hassler1990}, the constant nonthermal widths for all ions \citep{Tu1998}, or more complicated assumptions based on the nature of waves \citep[e.g.,][]{Dolla2008, Hahn2013b}. Besides, the line-of-sight (LOS) integration of the optically thin emission from the solar wind can also affect the observed line widths above the coronal hole \citep[e.g.,][]{Akinari2007,Gilly2020}.

In this study, we continued to study the dependence of $T_i$ on $Z/A$ in the polar coronal hole. To have better $Z/A$ coverage and a comparison between different instruments, we used a coordinated observation by SOHO/SUMER and Hinode/EIS. Because these two instruments cover different wavelengths and observe ions with different $Z/A$. We used the method proposed by \citet{Tu1998} to separate the thermal and nonthermal widths, which only assumes all ions have the same nonthermal velocity. Furthermore, we performed the global magnetohydrodynamic (MHD) simulation to validate the method and study the LOS integration effect. We describe the data reduction and analysis and MHD simulation in Section~\ref{sec:method}. Section~\ref{sec:result} shows the measured ion temperatures $T_i$ versus $Z/A$ and compares the observed and synthetic profiles. We discuss the factors affecting our diagnostics in Section~\ref{sec:dis}. Section~\ref{sec:conclusion} summarizes this study. 
\end{CJK*}

\section{Methodology} \label{sec:method}
\subsection{Observation and Data Reduction} \label{subsec:Obs}
\begin{figure*}[htb!]
    \centering
    \includegraphics[width=0.9\textwidth]{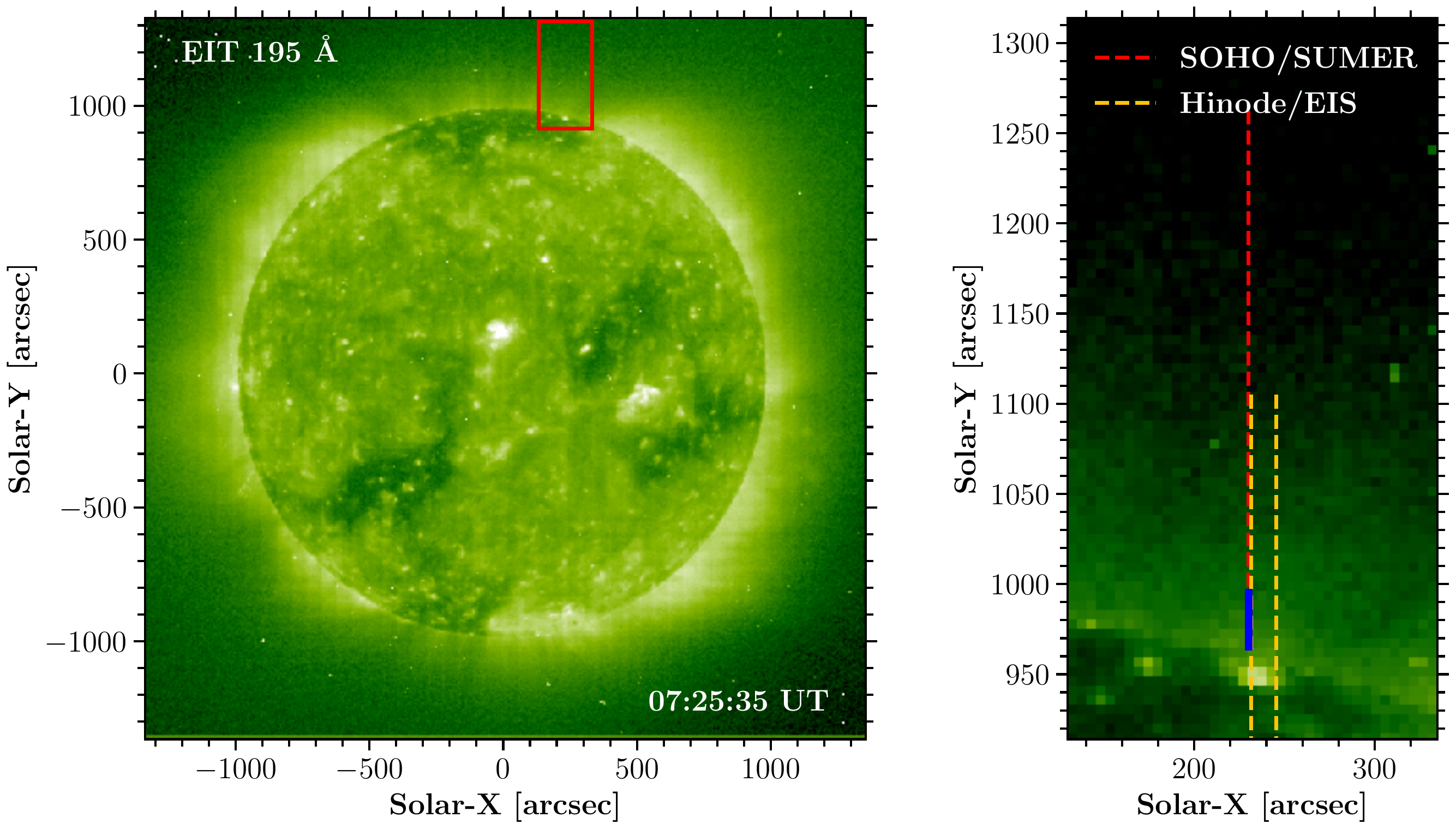}
    \caption{Positions of SUMER and EIS slits on the EIT 195 context image. \textbf{Left}: EIT 195 image of the full solar disk on 2007 November 16 at 07:25:35 UT. The red rectangle displays the FOV of the right panel. \textbf{Right}: slit pointing at the coronal hole boundary. The red dashed line shows the location of the $300\arcsec$ slit of SUMER. The yellow dashed lines show the first and the last pointing location of the EIS slit during the seven-step raster. The solid blue line shows the region of the data analyzed in this paper. Link to the \texttt{Jupyter} notebook creating this figure: \href{https://github.com/yjzhu-solar/EIS_SUMER_PCH_Ti/blob/main/ipynb/check_eit_img.ipynb}{\faGithub}.}
    \label{fig:eit}
\end{figure*}

On 2007 November 16, SOHO/SUMER and Hinode/EIS made a coordinated observation of the off-limb coronal hole boundary region at the north pole (see Figure~\ref{fig:eit}). SUMER observed this region from 09:01 UT to 10:03 UT. The center of the $4\arcsec\times300\arcsec$ SUMER slit 1 was pointed to $(230\arcsec,1120\arcsec)$, which covers the off-limb plasma from around 1.01 to 1.32\,$R_\odot$. SUMER detector B recorded the solar UV spectrum in four 45\,\mbox{\AA} wide spectrum windows. The four windows covered first-order wavelengths of 672\textendash 717\,\mbox{\AA}, 746\textendash 791\,\mbox{\AA}, 1015\textendash 1060\,\mbox{\AA}, and 1210\textendash 1255\,\mbox{\AA}. SUMER made three consecutive 300\,s exposures in each window. 

EIS observed this coronal hole from 07:26 UT to 08:01 UT in EUV from 170 to 210\,\mbox{\AA} (short wavelength, SW) and 245 to 290\,\mbox{\AA} (long wavelength, LW). The center of the EIS $2\arcsec\times 512\arcsec$ slit was pointed to $850\arcsec$ in Solar-Y. In the $x$ direction, EIS performed a seven-step raster scan from 232\farcs5 to 246\farcs5. The off-limb portion of the EIS slit covered the coronal hole plasma from about 1.00 to $1.15\,R_\odot$ (depending on the wavelength due to the tilt of the EIS grating and spatial offset between two EIS CCDs). EIS and SUMER have a spatial resolution of $1\arcsec$ along the slit. We note that \citet{Hahn2010} used the same EIS data set to study the ion temperature in the polar coronal hole, but we processed the EIS data with the latest EIS calibrations that were not available back then. 

We corrected and calibrated the SUMER and EIS data following the standard procedures described in the SUMER Data Cookbook \citep{SUMERDCBK} and the EIS software notes. We coaligned the SUMER and EIS observations using the EUV images taken by the Extreme-Ultraviolet Imaging Telescope \citep[EIT;][]{Delaboudiniere1995} on board the SOHO spacecraft (see Figure~\ref{fig:eit}). Details about the data calibration and coalignment can be found in Appendix~\ref{append:calib_coalign}. Finally, to maximize the number of observed ions, we averaged intensities of the 30 pixels (blue line in Figure~\ref{fig:eit}) between 1.01 and 1.04 solar radii for further analysis. Since this region is very close to the limb where the stray light does not significantly affect the fitting of the line width (see Appendix~\ref{appen:stray_los}), no stray-light correction is implemented to the EIS or SUMER data. 

\subsection{Data Fitting} \label{subsec:fit}

We performed the single/multi-Gaussian fitting to the unblended/blended spectral lines, using a constant background $I_{\rm bg}$. Each Gaussian profile is described by the total intensity $I_{\mathrm{tot},i}$, the wavelength of the line centroid $\lambda_{0,i}$, and the FWHM $\Delta \lambda_i$, i.e., the fitting model $I_{\rm model}$ can be written as
\begin{equation}
I_{\rm model}(\lambda) = \sqrt{\frac{4 \ln 2}{\pi}} \mathlarger{\sum}_{i=1}^{m} \frac{I_{\mathrm{tot},i}}{\Delta \lambda_i} \exp \left[- \frac{(\lambda - \lambda_{0,i})^2}{\Delta \lambda_i^2/4\ln 2} \right] + I_{\rm bg}
\end{equation}
where $m$ is the total number of spectral lines to be fitted. We used $\chi^2$ minimization to fit the complicated blended spectral lines.


To better estimate the fitting uncertainty in the single-Gaussian fitting, we adopted the Monte Carlo analysis \citep{Hahn2012} in the following steps: (1) We fit the unblended spectral line by simple $\chi^2$ minimization using the original uncertainty. (2) We reassigned the uncertainty of the intensity $\sigma_I$ to be the larger of the fitting residual or the original uncertainty. (3) We added noise to the observed intensities generated from a normal distribution $\mathcal{N}(\mu = 0,\sigma^2 = \sigma_I^2)$ and then fit the spectrum with additional noise. (4) We repeated step (3) 10,000 times and calculated the standard deviations of the fitting results as the fitting uncertainty.

We corrected the EIS line width according to the formula 
\begin{equation}
    \Delta \lambda_{\rm true} = \left(\Delta \lambda_{\rm fit}^2 - \Delta \lambda_{\rm inst}^2 \right)^{1/2}
\end{equation}
where $\Delta \lambda_{\rm fit}$ is the fitted FWHM and $\Delta \lambda_{\rm inst}$ is the EIS instrumental widths given by \citet{EISNote7}. The instrumental width $\Delta \lambda_{\rm inst}$ of SUMER is removed directly by the IDL routine \texttt{con\_width\_funct\_4} using deconvolution lookup tables based on measurements of \citet{Chae1998}.  

\subsection{Ion Temperature Estimation} \label{subsec:iontemp_meth}
The information about ion temperatures in the solar corona is embedded in spectral line widths $\Delta \lambda_{\rm true}$:
\begin{equation}
    \Delta \lambda_{\rm true} = \left[4\ln 2 \left(\frac{\lambda_0}{c} \right)^2 \left(\frac{2k_B T_i}{m_i} + \xi^2 \right)\right]^{1/2}
\end{equation}
where $c$ is the speed of light, $k_B$ is the Boltzmann constant, $T_i$ stands for the LOS ion temperature, $m_i$ is the mass of the ion, and $\xi$ is the nonthermal velocity. In observations, $\xi$ might be contributed by the propagation of Alfv\'en waves and other unresolved bulk motion along the LOS. As $\lambda_0$ is well determined, we can obtain the effective LOS velocity $v_{\rm eff}$ from the width of each fitted spectral line:
\begin{equation}
    v_{\rm eff} = \left(\frac{2k_B T_i}{m_i} + \xi^2 \right)^{1/2}
\end{equation}
The effective velocities $v_{\rm eff}$ of the lines originating from the same ion should be identical unless the lines of the same ion have very different excitation energies, which might cause the lines to form at different positions along the LOS. Therefore, to use multiple fitted lines from the same ion, we calculated the weighted average $\bar{v}_{\rm eff}$ of the effective velocity $v_{\rm eff}$ from $k$ spectral lines that have high signal-to-noise ratio (S/N) and small fitting uncertainties:
\begin{equation}
    \bar{v}_{\rm eff} = \frac{\sum_{i=1}^{k} w_i v_{\mathrm{eff},i}}{\sum_{i=1}^{k} w_i}
\end{equation}
where the weight is inversely proportional to the square of its uncertainty $w_i = 1/\sigma_{v,i}^2$. The uncertainty of the weighted average $\sigma_{\bar{v}}$ is given by \citet{Bevington2003}
\begin{equation}
    \sigma_{\bar{v}} = \sqrt{\frac{1}{k-1}\left( \frac{\sum_{i=1}^{k} w_i v_{\mathrm{eff},i}^2}{\sum_{i=1}^{k} w_i} - \bar{v}_{\rm eff}^2 \right)}
\end{equation}
In this way, multiple spectral lines provide a single $v_{\rm eff}$ measurement for each ion, which avoids confusion.

We estimated the limits of ion temperatures $T_i$ from the averaged effective velocities $\bar{v}_{\rm eff}$ using the method proposed by \citet{Tu1998}. This method assumes that the nonthermal velocity $\xi$ is the same for all ions. To estimate the upper limit, we assumed that the line profiles are dominated by thermal broadening, i.e., $\xi = 0$, therefore,
\begin{equation}
    T_{i,\mathrm{max}} = \frac{m_i}{2k_B}\bar{v}_{\rm eff}^2
\end{equation}
To obtain the lower limit $T_{i,\mathrm{min}}$, we first set the upper limit of the nonthermal velocity $\xi_{\rm max}$ to be the smallest effective velocity $\bar{v}_{\rm eff}$ among all lines. Then, we removed $\xi_{\rm max}$ from all $\bar{v}_{\rm eff}$ to calculate $T_{i,\mathrm{min}}$
\begin{equation}
    T_{i,\mathrm{min}} = \frac{m_i}{2k_B}\left(\bar{v}_{\rm eff}^2 - \xi_{\rm max}^2\right)
\end{equation}
We note that the interval $[T_{i,\mathrm{min}},T_{i,\mathrm{max}}]$ should not be interpreted as an uncertainty but as a range of equally likely values. We propagated the uncertainty in the average effective velocity $\bar{v}_{\rm eff}$ to the uncertainty of each $T_{i,\mathrm{max}}$ and $T_{i,\mathrm{min}}$.

Given the uncertainties in the instrumental width, we carried out the $T_i$ diagnostics for SUMER and EIS lines separately, so that the ions observed by the two instruments had an EIS measurement and a SUMER measurement. 

\subsection{Electron Density and Temperature Diagnostics} \label{subsec:electemp}

We measured the electron density $n_e$ and temperature $T_e$ at the coronal hole boundary using the intensity ratios of the two lines originating from the same ion. The measured $T_e$ would be compared with $T_i$ to study the preferential heating of heavy ions. We listed the lines used for $n_e$ and $T_e$ diagnostics in Tables~\ref{tab:lineNe} and \ref{tab:lineTe}. We assumed that the plasma along the LOS is nearly homogeneous in density and isothermal and used the latest CHIANTI~10 atomic database \citep{Dere1997,DelZanna2021}.

\begin{figure}[htb!]
    \centering
    \includegraphics[width=\linewidth]{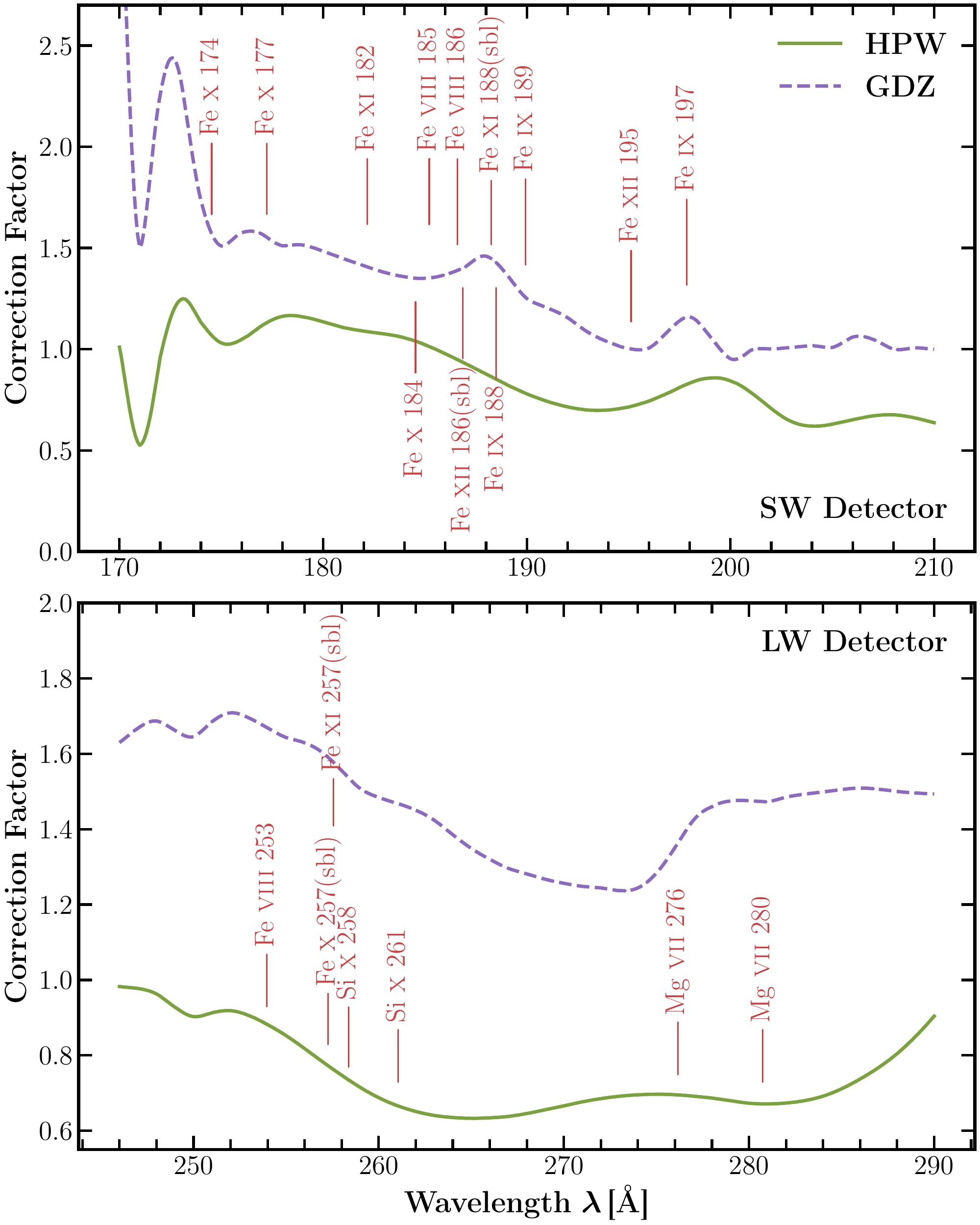}
    \caption{Two in-flight radiometric corrections in the SW (\textbf{top}) and LW (\textbf{bottom}) detectors of EIS. Locations of the spectral lines used for electron temperature and density diagnostics are also displayed. Link to the \texttt{Jupyter} notebook creating this figure: \href{https://github.com/yjzhu-solar/EIS_SUMER_PCH_Ti/blob/main/ipynb/eis_recalibrate_comp.ipynb}{\faGithub}.}
    \label{fig:radiometry}
\end{figure}

Besides the fitting uncertainty, the radiometric calibration might also affect the precision of electron density and temperature diagnostics. The SUMER radiometric calibration has an absolute uncertainty of 36\% in detector B \citep{Pauluhn2001} after the SOHO recovery in 1998. Although the ratios of two lines might have a smaller uncertainty, we conservatively use those values. EIS has two competing in-flight radiometric corrections to the laboratory calibration proposed by \citet{DelZanna2013} (GDZ) and \citet{Warren2014} (HPW). The two methods show different detector responses in the SW and LW detectors. We show the variation of two correction factors versus wavelength and the wavelengths of the spectral lines used for diagnostics in Figure~\ref{fig:radiometry}. The two correction curves are similar, but the GDZ curve has more small bumps. For example, the increase in the GDZ correction factor around 188\,\mbox{\AA} makes the Fe \textsc{viii} 185.213/186.598\,\mbox{\AA} ratio 10\% smaller than the HPW ratio. The difference in the absolute LW detector calibration affects most EIS ratios used to measure the electron temperature. 

We applied the two corrections to EIS line intensities to investigate their differences. The exception is the Fe \textsc{ix} 188/189\,\mbox{\AA} ratio because the two lines are very close on the detector. Another exception is the Mg \textsc{ix} 706\,\mbox{\AA} line observed by SUMER, located at the boundary of the coated and bare part of the detector. Although the KBr and bare responsivities at $\sim 700$\,\mbox{\AA} are very similar, we still measured the electron density and temperature assuming that the entire Mg~\textsc{ix} 706\,\mbox{\AA} line is recorded in the KBr or bare parts.

\begin{deluxetable*}{cccc}

\tablecaption{Spectral lines ratios used for electron density diagnostics.}
\label{tab:lineNe}

\tablehead{\colhead{Ion} & \colhead{$\log T_{\rm max}$ (K)} & \colhead{Wavelength (\mbox{\AA})} & \colhead{Instrument}} 

\startdata
Fe \textsc{viii} & 5.75 & 185.213/186.598 & EIS \\
Mg \textsc{vii} & 5.80 & 276.154/280.742 & EIS \\
Fe \textsc{ix} & 5.85 & 188.493/189.935 & EIS \\
Mg \textsc{ix} & 5.95 & 694.006/706.060 & SUMER \\
Si \textsc{x} & 6.00 & 258.374/261.056 & EIS \\
Fe \textsc{xi} & 6.00 & 182.167/(188.216+188.299) & EIS \\
Fe \textsc{xii} & 6.05 & (186.854+186.887)/195.119 &  EIS \\
\enddata

\tablecomments{$T_{\rm max}$ is the temperature of the maximum line formation calculated by CHIANTI using the DEM derived from the average coronal hole spectra in \citet{Vernazza1978}}.

\end{deluxetable*}

\begin{deluxetable*}{cccc}




\tablecaption{Spectral lines ratios used for electron temperature diagnostics.}
\label{tab:lineTe}


\tablehead{\colhead{Ion} & \colhead{$\log T_{\rm max}$ (K)} & \colhead{Wavelength (\mbox{\AA})} & \colhead{Instrument}} 

\startdata
Fe \textsc{viii} & 5.75 & 185.213/253.956 & EIS \\
Fe \textsc{ix} & 5.85 & 191.206/197.854 & EIS \\
Fe \textsc{x} & 5.95 & 174.531/(257.259+257.261) & EIS \\
Fe \textsc{x} & 5.95 & 177.240/(257.259+257.261) & EIS \\
Fe \textsc{x} & 5.95 & 184.537/(257.259+257.261) & EIS \\
Mg \textsc{ix} & 5.95 & 706.060/749.552 & SUMER \\
Fe \textsc{xi} & 6.00 & (188.216+188.299)/(257.547+257.554) & EIS \\
\enddata
\tablecomments{$T_{\rm max}$ is the temperature of the maximum line formation calculated by CHIANTI using the DEM derived from the average coronal hole spectra in \citet{Vernazza1978}. }



\end{deluxetable*}

\subsection{AWSoM-R Simulation} \label{subsec:awsom}
We used the MHD simulation to study the LOS integration effect, including the line formation regions and macroscopic Doppler broadening due to the bulk motions. We simulated the global solar corona and synthesized line profiles at the coronal hole boundary with the Space Weather Modeling Framework \citep[SWMF;][]{Toth:2012} Alfv\'en Wave Solar Model - realtime  \citep[AWSoM-R;][]{Sokolov:2021} model combined with the postprocessing tool SPECTRUM \citep{Szente:2019}, which is also part of the SWMF. AWSoM-R avoids the artificial stretching of the transition region \citep{Lionello:2001} in the Alfv\'en Wave Solar Model \citep[AWSoM;][]{vanderHolst:2014} using the threaded-field-line model, which provides a better comparison with observations close to the limb. For more details about AWSoM-R, we refer the readers to \citet{Sokolov:2021}.


The SPECTRUM module calculates the emissivity profiles, including Doppler shift and line broadening, voxel by voxel, and integrates them along the LOS. The FWHM of the emissivity profile is 
\begin{eqnarray}
  \label{spe:linebroadening}
      \Delta \lambda = \left[4\ln 2 \left(\frac{\lambda_0}{c} \right)^2 \left(\frac{2 k_B T_{p}}{ m_{p} A_{i}}  + \xi_w^2 \right) \right]^{1/2}
  \end{eqnarray} 
  where $m_p$ is the proton mass and $A_i$ is the mass number of the ion. AWSoM-R does not model heavy-ion species, so we assumed that the ion temperature $T_i$ equals proton temperature $T_p$. The wave-induced nonthermal velocity $\xi_w$ is 
\begin{eqnarray}
  \label{spe:eq:vnonthermal}
  \xi_w^2 = \frac{1}{2} \langle \delta u^2 \rangle\sin^2{\alpha} =
  \frac{1}{2}\frac{\omega^{+}+\omega^{-}}{\rho} \sin^2{\alpha} =  \nonumber \\
  \frac{1}{8}  \left(z_{+}^2 + z_{-}^2 \right)\sin^2{\alpha} .  
\end{eqnarray}
where $\alpha$ is the angle between the direction of the local magnetic field and the LOS, and $z_{\pm}$ are the Els\"{a}sser variables for forward- and backward-propagating waves and the respective energy densities are $\omega_{\pm}$. More information about SPECTRUM can be found in \citet{Szente:2019}.


\section{Results} \label{sec:result}
\subsection{Observations}

\subsubsection{Line Fitting and Average Effective Velocity}
We fitted all available SUMER and EIS spectral lines. Appendix~\ref{appen:fit_example} shows examples of single and multi-Gaussian fittings. The EIS and SUMER line widths used for ion temperature diagnostics are listed in Table~\ref{tab:line_width}. 

\begin{figure}
    \centering
    \includegraphics[width=0.95\linewidth]{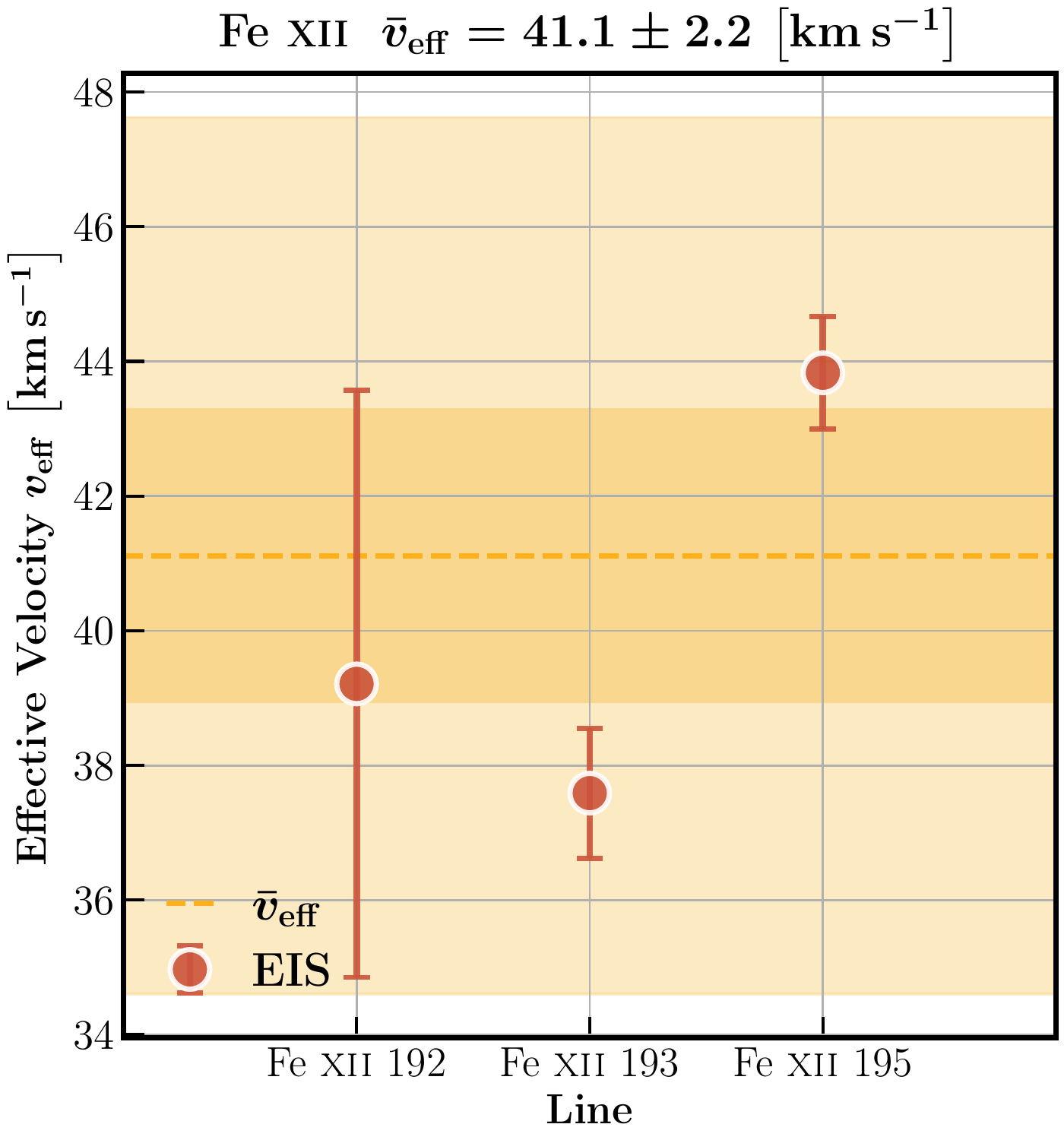}
    \caption{The measured effective velocity $v_{\rm eff}$ of the Fe \textsc{xii} 192, 193, and 195 \mbox{\AA} lines and the average effective velocity $\bar{v}_{\rm eff}$ of Fe \textsc{xii}. The horizontal dashed line indicates the average effective velocity $\bar{v}_{\rm eff}$. The darker/lighter shaded regions show the $1\sigma$/$3\sigma$ uncertainty level, respectively. Link to the \texttt{Jupyter} notebook creating this figure: \href{https://github.com/yjzhu-solar/EIS_SUMER_PCH_Ti/blob/main/ipynb/sim_obs_comp/eis_average_eff.ipynb}{\faGithub}.} 
    \label{fig:FeXII_bootstrap}
\end{figure}

Then we calculated the average effective velocity for ions with multiple spectral lines observed by the same instrument. We separately treated spectral lines originating from the same ion but observed by different instruments for comparison (e.g., the O \textsc{vi} 184\,\mbox{\AA} line observed by EIS and the O \textsc{vi} 1032, 1037\,\mbox{\AA} lines observed by SUMER). As an example, Figure~\ref{fig:FeXII_bootstrap} shows the effective velocity of the Fe \textsc{xii} 192.394, 193.509, and 195.119\,\mbox{\AA} triplet lines and the average effective velocity $\bar{v}_{\rm eff}$. The Fe \textsc{xii} 192\,\mbox{\AA} line width shows greater uncertainty because of the low intensity and blending. The Fe \textsc{xii} 195\,\mbox{\AA} line is slightly broader than the 192 and 193\,\mbox{\AA} lines by $\sim 5\,\mathrm{km\,s^{-1}}$, which could be related to the instrumental effect \citep{DelZanna2019} and the blended Fe \textsc{xii} 195.179\,\mbox{\AA} line \citep{Young2009}. The average effective velocity of Fe \textsc{xii} is $41.1\pm2.2\,\mathrm{km\, s^{-1}}$. 

\subsubsection{Electron Density and Temperature Diagnostics}\label{subsubsec:Te_Ne_res}
\begin{figure*}[htb!]
    \centering
    \includegraphics[width=\linewidth]{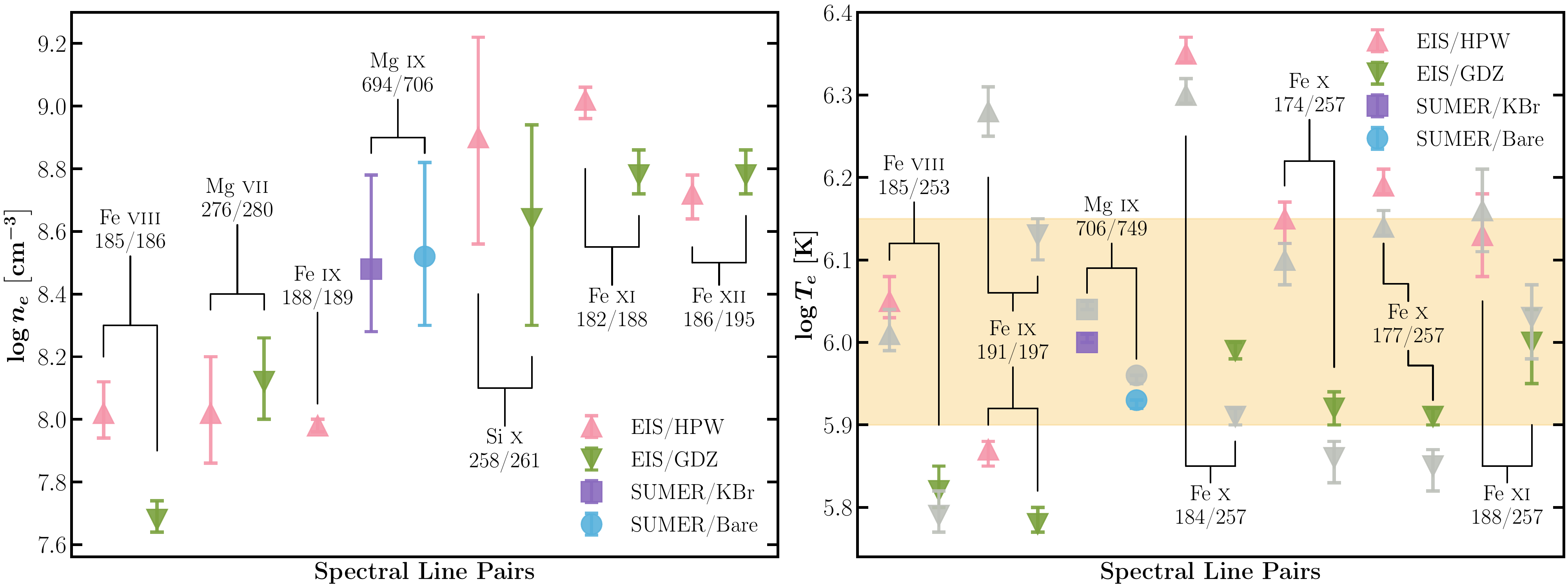}
    \caption{\textbf{Left:} electron density $n_e$ diagnostics of the coronal hole boundary region shown in Figure~\ref{fig:eit}. \textbf{Right:} electron temperature $T_e$ diagnostics in the same region. The colored data points in the right panel represent $T_e$ at $\log n_e = 8.0$, while the gray ones stand for the $T_e$ inferred at $\log n_e = 8.5$. The yellow-shaded area displays the chosen range of the electron temperature. The line ratios are sorted from left to right by the maximum formation temperature $T_{\rm max}$ (see Tables~\ref{tab:lineNe} and \ref{tab:lineTe}). Link to the \texttt{Jupyter} notebook creating this figure: \href{https://github.com/yjzhu-solar/EIS_SUMER_PCH_Ti/blob/main/ipynb/paper/Te_Ne_diag.ipynb}{\faGithub}. }
    \label{fig:Ne_Te}
\end{figure*}

Figure~\ref{fig:Ne_Te} summarizes the measured $n_e$ and $T_e$, where the ions are ordered by their formation temperature. The measured electron densities range from $\log n_e \sim 7.7$ to $\log n_e \sim 9.0$, depending on the line pairs and calibration methods. The measured electron density increases with the maximum formation temperature of the ion. The electron densities measured from cooler ($\log T_{\rm max} < 5.9$) line pairs like Fe \textsc{viii} 185/186\,\mbox{\AA}, Mg \textsc{vii} 276/280\,\mbox{\AA}, and Fe \textsc{ix} 188/189\,\mbox{\AA} are around $\log n_e \sim 8.0$. The hotter Mg \textsc{ix} 694/706\,\mbox{\AA} ($\log T_{\max} \sim 5.95$) ratio gives the electron density $\log n_e \sim 8.5$. We found the highest electron densities $\log n_e \sim 8.7$\textendash9.0 in the hottest ($\log T_{\rm max} > 6.0$) Si \textsc{x} 258/261\,\mbox{\AA}, Fe \textsc{xi} 182/188\,\mbox{\AA}, and Fe \textsc{xii} 186/195\,\mbox{\AA} lines. The increase in $n_e$ with $T_{\rm max}$ implies that off-limb emissions might have two different source regions along the LOS: a cooler and less dense coronal hole and a hotter and denser region. However, it is difficult to identify the hotter region from both the observation and the simulation in Section~\ref{subsec:simulations}. 

We chose two electron densities $\log n_e = 8.0$ and $\log n_e = 8.5$ to calculate temperature-sensitive line ratios using CHIANTI. The inferred electron temperatures from different line ratios are shown in the right panel of Figure~\ref{fig:Ne_Te}. Most electron temperatures range from $\log T_e \sim 5.9$ to $\log T_e \sim 6.2$. EIS electron temperatures show a wider distribution due to different radiometric calibrations. The $T_e$ inferred from the HPW correction are higher than the $T_e$ inferred from the GDZ correction by 0.1\textendash 0.3 dex (a factor of 1.3\textendash 2.0 on the linear scale). The $T_e$ measured from hotter line pairs like Fe \textsc{x} 174/257 and Fe \textsc{xi} 188/257 are slightly greater than $T_e$ inferred from cooler line ratios like Fe \textsc{viii}. Although the Fe \textsc{ix} 191/197\,\mbox{\AA} ratio is less affected by the cross-calibration between two detectors, the line ratio is more sensitive to the electron density. We measured a higher $T_e$ in the Fe \textsc{x} 184/257\,\mbox{\AA} ratio than the other two Fe \textsc{x} 174/257\,\mbox{\AA} and 177/257\,\mbox{\AA} ratios. As the magnetic field in the coronal hole is very weak, the blended Fe \textsc{x} 257\,\mbox{\AA} magnetic-induced transition (MIT) should not affect the Fe \textsc{x} line ratios. The $T_e$ inferred from the Mg \textsc{ix} 706/749\,\mbox{\AA} ratio lies between most of the HPW and GDZ electron temperatures. We chose $\log T_e = 5.9$ to $\log T_e = 6.15$ as the electron temperature of the LOS plasma to cover most of the measurements.  

\subsubsection{Ion Temperature Diagnostics}\label{subsubsec:Ti_diag}

Although the $n_e$ measurements suggest the presence of two different structures along the LOS, we found it challenging to perform $T_i$ diagnostics on each structure separately. First, the two regions might contribute emissivity to all observed spectral lines, although with different weights. Second, the number of ions is too limited to carry out $T_i$ diagnostics in each region separately. Third, it would be more complicated to treat lines like Fe \textsc{xii} and Fe \textsc{xiii}, whose widths might be affected by the bulk motions along the LOS (see Section~\ref{subsec:simulations}). Hence, we continued to use all spectral lines to measure $T_i$. Note that the measured $T_i$, like other measurements in the optically thin plasma, is an average along the LOS. 

\begin{figure*}[htb!]
    \centering
    \includegraphics[width=\linewidth]{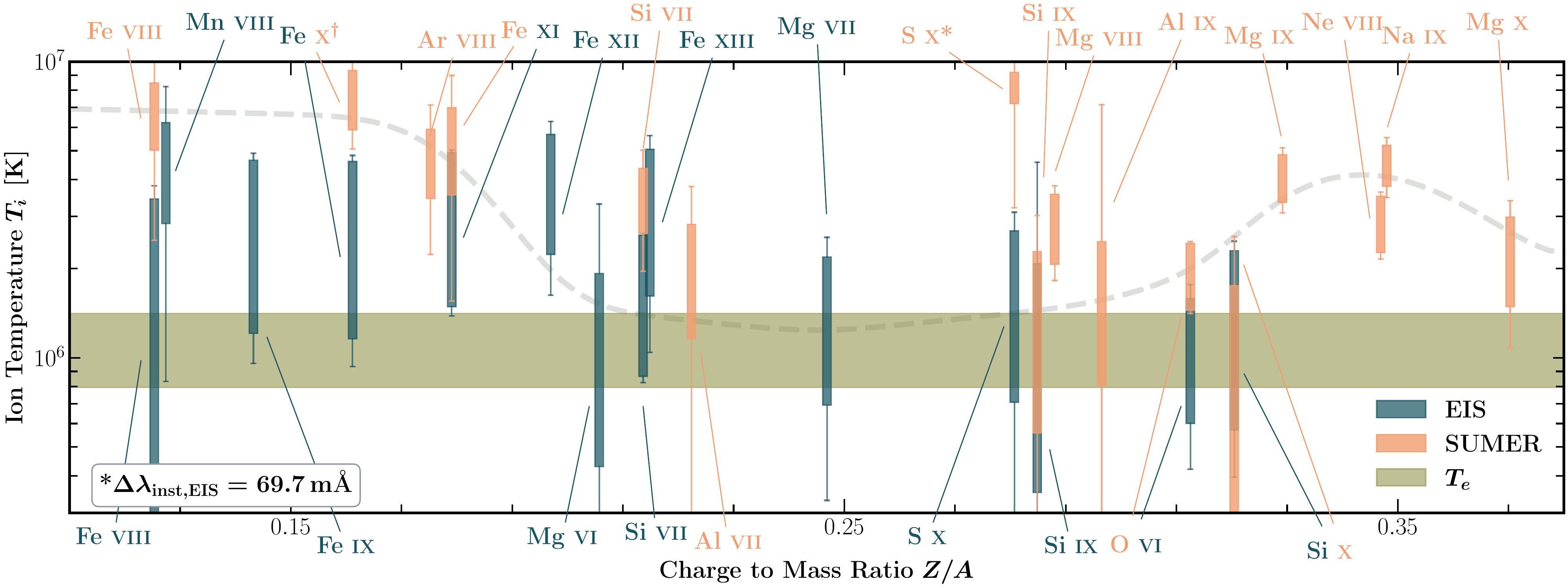}
    \caption{Estimated ion temperatures $T_i$ vs. ion charge-to-mass ratio $Z/A$. For each ion, the boxes show possible ion temperature intervals between $T_{i,\mathrm{min}}$ and $T_{i,\mathrm{max}}$, while the whiskers indicate the uncertainty in $T_{i,\mathrm{min}}$ and $T_{i,\mathrm{max}}$. The ion temperatures inferred from lines observed by EIS and SUMER are shown in dark green and orange, respectively. The horizontal shaded area displays the electron temperature measurement shown in Figure~\ref{fig:Ne_Te}. The dashed gray line is arbitrarily drawn to illustrate the U-shaped dependence. $^*$ The S \textsc{x} 776.373\,\mbox{\AA} line is poorly fitted. $^\dagger$The Fe \textsc{x} 1028\,\mbox{\AA} line is self-blended, but still kept for comparison. Link to the \texttt{Jupyter} notebook creating this figure: \href{https://github.com/yjzhu-solar/EIS_SUMER_PCH_Ti/blob/main/ipynb/paper/temp_diag_v2.ipynb}{\faGithub}.}
    \label{fig:temp_diag}
\end{figure*}

Figure~\ref{fig:temp_diag} shows the minimum and maximum of the ion temperature $T_{i,\mathrm{min}}$ and $T_{i,\mathrm{max}}$ versus the ion $Z/A$, along with the $T_e$ determined in Figure~\ref{fig:Ne_Te}. The values of $T_{i,\mathrm{min}}$ and $T_{i,\mathrm{max}}$ are also listed in Table~\ref{tab:line_width}. We measured the $T_i$ of ions with $Z/A$ ranging from 0.125 (Fe \textsc{viii}) to 0.37 (Mg \textsc{x}). Since the coronal hole plasma is cooler than the plasma in the quiet-Sun or active regions, spectral lines from higher charge states are too weak to identify or fit. The narrowest lines are from Fe \textsc{viii} (EIS) and Si \textsc{x} (SUMER), with similar effective velocities $\bar{v}_{\rm eff} \sim 32\,\mathrm{km\, s^{-1}}$. We used the $\bar{v}_{\rm eff}$ of Fe \textsc{viii} as the maximum nonthermal velocity $\xi_{\rm max}$. Note that the Si \textsc{x} 624.694\,\mbox{\AA} line observed by SUMER is blended with the stronger Mg \textsc{x} 624.941\,\mbox{\AA} line, so the uncertainty of $v_{\rm eff}$ is quite large. 

The measured $T_i$ shows a U-shaped dependence on the charge-to-mass ratio $Z/A$ similar to the $T_i$ measured by \citet{Landi2009} using a SUMER observation at the center of the coronal hole. Ions with $0.12 \leq Z/A \leq 0.19$, except for the two ions Fe \textsc{viii} (EIS) and Fe \textsc{ix} with the smallest $v_{\rm eff}$, show $T_i$ much higher than the local $T_e$.  The $T_i$ decreases with $Z/A$ between 0.19 and 0.25, then shows a plateau close to the local $T_e$ ranging from $Z/A = 0.25$ to 0.33. Above $Z/A = 0.33$, the $T_i$ becomes greater than the local $T_e$ again. Mg \textsc{x} with the greatest $Z/A = 0.37$ reveals a lower $T_i$ than the other ions with $Z/A > 0.33$ (Mg \textsc{ix}, Ne \textsc{viii} and Na \textsc{ix}). 

The observation did not suggest a clear dependence of $T_i$ on formation temperatures $T_{\rm max}$. Although some cooler ions observed by EIS, like Mg \textsc{vii} and O \textsc{vi}, show lower $T_i$ compared to the hotter ions like Fe \textsc{xii}, the $T_i$ of ions with similar $T_{\rm max}$ can be dramatically different, e.g., Fe \textsc{viii} (SUMER) and Mg \textsc{vii} (EIS). We further investigate the influence of the ion formation temperature and bulk motions along the LOS in the AWSoM-R simulation (see Section~\ref{subsec:simulations}).

Contrary to expectations, in most cases, the $\bar{v}_{\rm eff}$ and the $T_i$ measured by SUMER are greater than $\bar{v}_{\rm eff}$ and $T_i$ from the same ion measured by EIS. The only exception is Si \textsc{x}, which might be due to the fitting uncertainty caused by line blending. Both EIS and SUMER record strong emission lines from Fe \textsc{viii} and O \textsc{vi}, providing the most reliable comparisons. The $v_{\rm eff} = 50.2 \pm 7.5\,\mathrm{km \, s^{-1}}$ of the SUMER Fe \textsc{viii} is 30\% to 50\% higher than the EIS value of $\bar{v}_{\rm eff} = 32.0\pm1.7\,\mathrm{km \, s^{-1}}$, while the SUMER O \textsc{vi} $\bar{v}_{\rm eff} = 50.3 \pm 0.6\,\mathrm{km \, s^{-1}}$ is also about 25\% larger than the EIS value of $v_{\rm eff} = 40.6 \pm 2.3\,\mathrm{km \, s^{-1}}$. We will discuss the uncertainty caused by instrumental broadening in Section~\ref{subsec:instbroad}.

\subsection{Simulations}\label{subsec:simulations}
\begin{figure*}[htb!]
    \centering
    \includegraphics[width=\linewidth
    ]{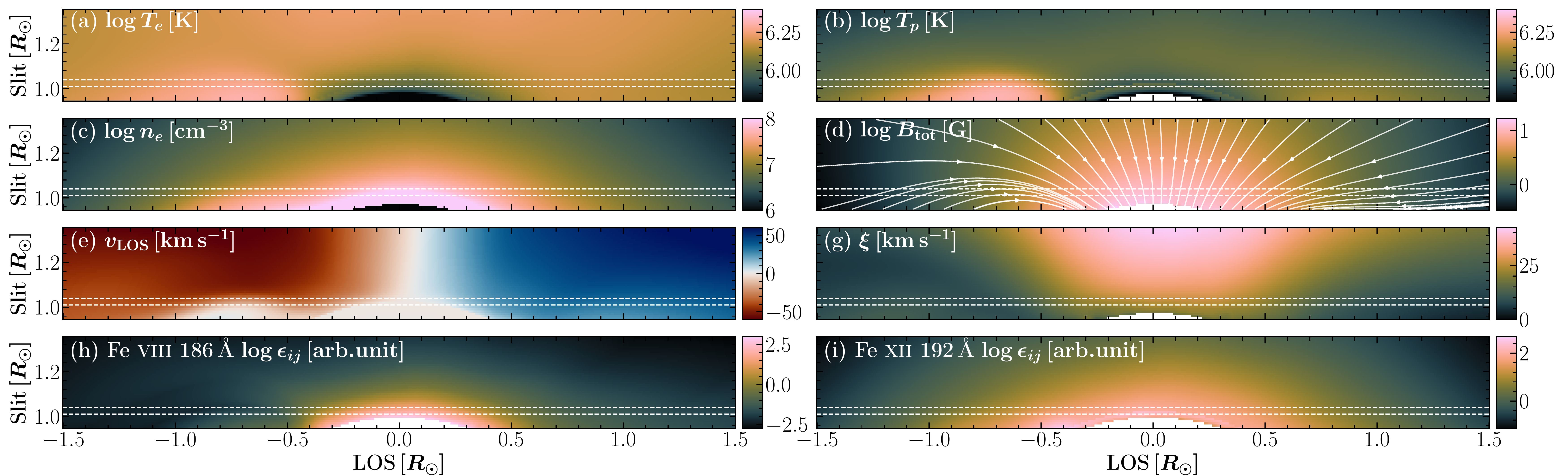}
    \caption{Physical parameters predicted by AWSoM-R along the LOS covering the entire SUMER slit. The dotted white lines show the region selected for analysis. \textbf{(a)} Electron temperature $T_e$, \textbf{(b)} proton temperature $T_p$, \textbf{(c)} electron density $n_e$, \textbf{(d)} total magnetic field $B_{\rm tot}$ and field lines, \textbf{(e)} LOS velocity $v_{\rm LOS}$,\textbf{(g)} wave-induced nonthermal velocity $\xi_w$, \textbf{(h)} emissivity of the Fe \textsc{viii} 186\,\mbox{\AA} line, \textbf{(i)} and emissivity of the Fe \textsc{xii} 192\,\mbox{\AA} line. Link to the \texttt{Jupyter} notebook creating this figure: \href{https://github.com/yjzhu-solar/EIS_SUMER_PCH_Ti/blob/main/ipynb/awsom_los/awsomr_100k_los.ipynb}{\faGithub}.}
    \label{fig:awsom_los}
\end{figure*}

Figure~\ref{fig:awsom_los} shows the physical parameters in the AWSoM-R simulation. The coronal hole boundary region in the simulation does not show complicated structures along the LOS, except for a streamer at the far side. The $T_e$ between 1.01 and 1.04\,$R_\odot$ is $\sim 1\,$ MK, and the electron density $n_e$ is $\sim 10^8\,\mathrm{cm^{-3}}$, matching the diagnostics results  in Section~\ref{subsubsec:Te_Ne_res}. The LOS velocity varies from 0 to $\pm 20\mathrm{km \, s^{-1}}$ where most spectral lines form in the studied region. The wave-induced nonthermal velocity $\xi$ is about $20\,\mathrm{km\, s^{-1}}$ in the open field lines between 1.01 and 1.04\,$R_\odot$. 

The maximum formation temperature $T_{\rm max}$ of the spectral line affects the line formation region along the LOS in Figure~\ref{fig:awsom_los}. For example, most Fe \textsc{viii} 186\,\mbox{\AA} emissions ($\log T_{\rm max} \sim 5.75$) are contributed by the plasma from $- 0.3\,R_\odot$ to $0.3\,R_\odot$ along the LOS. Most of the hotter Fe \textsc{xii} 192\,\mbox{\AA} ($\log T_{\rm max} \sim 6.05$) emission forms between $-0.5\,R_\odot$ and $0.5\,R_\odot$. The streamer at the far side does not contribute enough emissions to Fe \textsc{xii} to influence the density diagnostics. We did not find other hotter and denser regions along the LOS in the simulation. 

\begin{figure*}
    \centering
    \includegraphics[width=\linewidth]{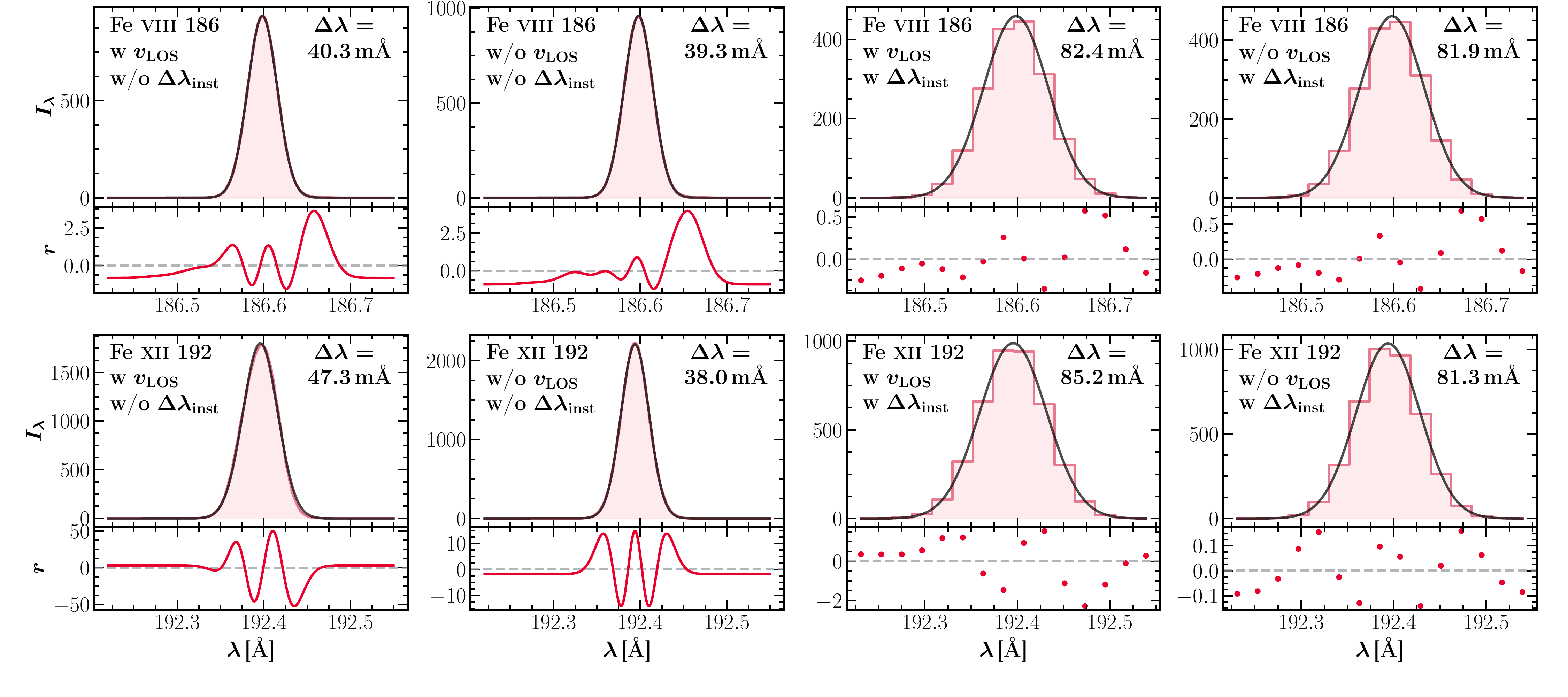}
    \caption{Fitting synthetic Fe \textsc{viii} 186\,\mbox{\AA} and Fe \textsc{xii} 192\,\mbox{\AA} lines under different conditions. w $v_{\rm LOS}$: synthetic profiles with the Doppler effect. w/o $v_{\rm LOS}$: synthetic profiles without the Doppler effect. w $\Delta \lambda_{\rm inst}$: synthetic profiles convolved with instrumental width $\Delta \lambda_{\rm inst} = 70\,$m\mbox{\AA} and rebinned to EIS spectral resolution. w/o $\Delta \lambda_{\rm inst}$: synthetic profiles without instrumental effects. Link to the \texttt{Jupyter} notebook creating this figure: \href{https://github.com/yjzhu-solar/EIS_SUMER_PCH_Ti/blob/main/ipynb/spectrum_fit/DopplerVsNoDoppler.ipynb}{\faGithub}.}
    \label{fig:awsom_vLOS}
\end{figure*}

To evaluate the influence of LOS bulk motions on line profiles, we synthesized Fe \textsc{viii} 186\,\mbox{\AA} and Fe \textsc{xii} 192\,\mbox{\AA} profiles either with or without the Doppler effect in Figure~\ref{fig:awsom_vLOS}. The macroscopic Doppler broadening only increases the Fe \textsc{viii} width by about 2.5\%. The LOS integration of bulk motions in Fe \textsc{viii} becomes even more negligible when the profile is convolved with the instrumental width and degraded to EIS spectral resolution. The actual width of the hotter Fe \textsc{xii} 192\,\mbox{\AA} line increases from 38.0 to 47.3\,\mbox{\AA} due to the macroscopic Doppler broadening. After the convolution with the instrumental widths, the bulk motion still increases the Fe \textsc{xii} line width from 81.3 to 85.2\,\mbox{\AA}, suggesting that the bulk motions might increase line widths noticeably in actual observations.    

\begin{figure*}
    \centering
    \includegraphics[width=\linewidth]{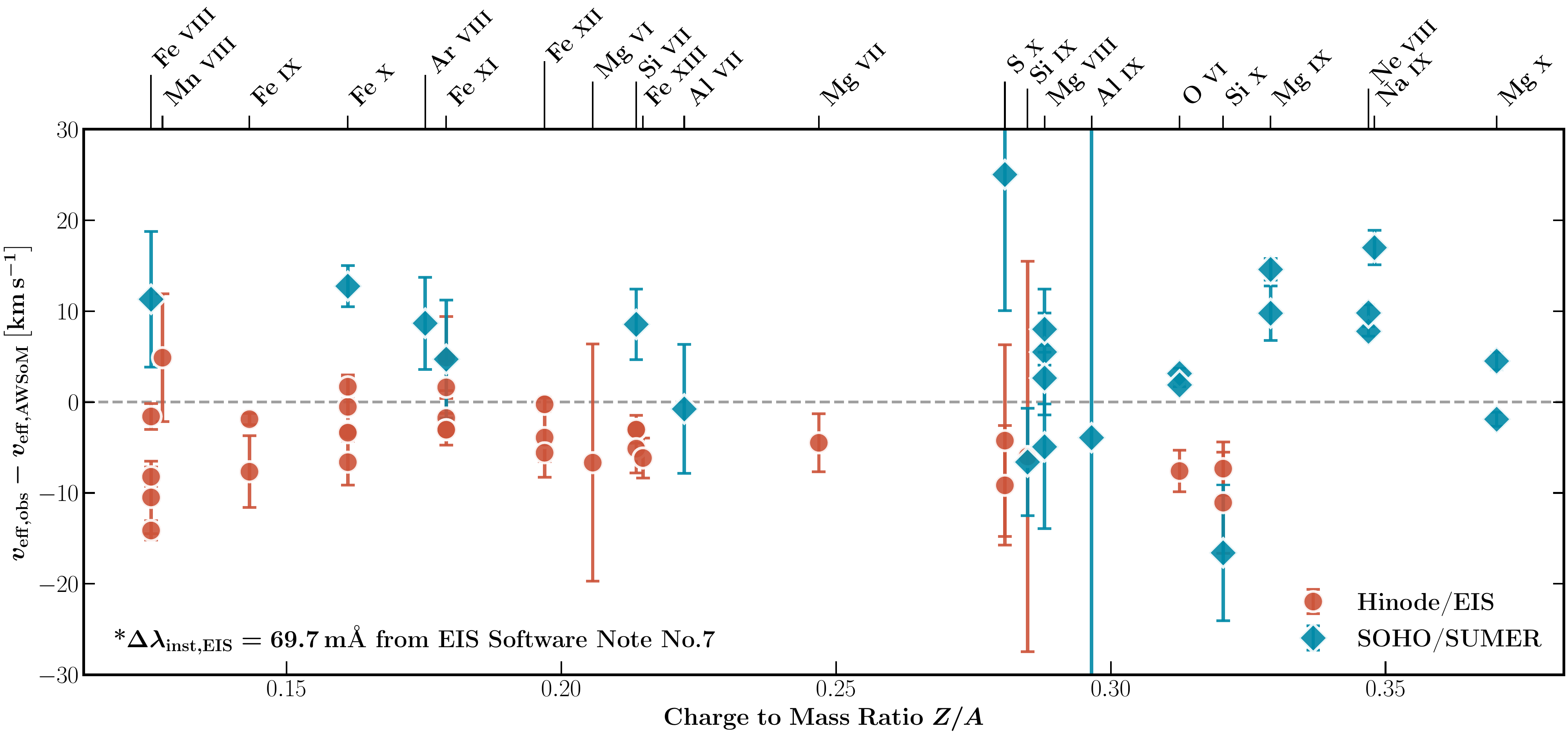}
    \caption{Differences between the observed effective velocity $v_{\rm eff, obs}$ and the synthetic effective velocity $v_{\rm eff, AWSoM}$ of each spectral line used in ion temperature diagnostics. Red dots with error bars: EIS lines. Blue diamonds with error bars: SUMER lines. Link to the \texttt{Jupyter} notebook creating this figure:  \href{https://github.com/yjzhu-solar/EIS_SUMER_PCH_Ti/blob/main/ipynb/sim_obs_comp/sim_obs_linewidth.ipynb}{\faGithub}.}
    \label{fig:awsom_obs}
\end{figure*}

Figure~\ref{fig:awsom_obs} compares the observed line widths with the synthetic line widths. The SPECTRUM module uses the proton temperature $T_p$ to approximate the $T_i$ to calculate the thermal broadening. Therefore, the excessive widths in the observation might indicate the additional heating in heavy ions compared to the proton. Most of the spectral lines observed by SUMER show widths similar to or larger than the synthetic widths. On the other hand, most of the EIS lines are $\sim$5\textendash10\,$\mathrm{km\,s^{-1}}$ narrower than the synthetic lines. The differences between the SUMER and synthetic line widths show a similar U-shaped dependence on $Z/A$, with SUMER lines being wider than synthetic ones by $\sim$5\textendash 20\,$\mathrm{km\,s^{-1}}$. For ions observed by both EIS and SUMER, AWSoM-R underestimates the SUMER line widths but overestimates the EIS line widths (e.g., Fe \textsc{viii}, Fe \textsc{xi}, and O \textsc{vi}). This is because the $v_{\rm eff}$ measured by SUMER is usually greater than the $v_{\rm eff}$ of the same ion observed by EIS (see Section~\ref{subsubsec:Ti_diag}). 

\section{Discussion} \label{sec:dis}
\subsection{Uncertainty in Instrumental Broadening} \label{subsec:instbroad}

Inconsistency in the $v_{\rm eff}$ and $T_i$ of the same ion observed by SUMER and EIS might result from the instrumental widths used in this study. The instrumental widths of the EIS $2\arcsec$ slit are measured by searching for the smallest Fe \textsc{xii} 193.509\,\mbox{\AA} line widths in a series of off-limb quiet-Sun observations. Only a thermal width of $\Delta \lambda_{\rm th} = 23.2\,$m\mbox{\AA} was removed from the Fe \textsc{xii} width to obtain the instrumental width as a function of position along the slit, assuming $\log T_i =\log T_{\rm max} = 6.2$ \citep{EISNote7}. This raises the concern that the EIS instrumental width might be overestimated; also, the instrumental widths might depend on the wavelength (see the discussion in Appendix~\ref{append:dlamb_EIS_lamb}).  At the averaged 30 pixels, the EIS instrumental width for the $2\arcsec$ slit is $69.7$\,m\mbox{\AA}, with an uncertainty of $\sim 3$\,m\mbox{\AA}. 

The SUMER instrumental widths were determined using the widths of narrow neutral lines in quiet-Sun observations \citep{Chae1998} and P. Lemaire's calculations of 1997 August 28. \citet{Chae1998} reported the instrumental width of SUMER detector B with 1\arcsec slit is about 3.0 pixels ($\approx129$\,m\mbox{\AA} in the first order) with a fluctuation of 0.5 pixels ($\approx22$\,m\mbox{\AA}). Compared with the EIS instrumental width ($69.7$\,m\mbox{\AA}), which dominates the observed profiles of 80\textendash90\,m\mbox{\AA}, the SUMER instrumental width of $\sim 129$\,m\mbox{\AA} contributes much less to the observed widths between 200 and 350\,m\mbox{\AA} at the first order. 

To investigate the influence of instrumental widths on the diagnostic results, we cross-calibrated the EIS instrumental widths $\Delta \lambda_{\rm eff, EIS}$ with the SUMER observation by matching the width of O \textsc{vi} 184.117\,\mbox{\AA} (EIS) to the widths of O \textsc{vi} 1031.912 and 1037.613\,\mbox{\AA} (SUMER). We assumed that SUMER correctly measured the O \textsc{vi} width. That is because the SUMER in-flight instrumental widths are measured from the narrowest neutral lines, where the thermal and nonthermal broadening are negligible. We obtained a new EIS instrumental width $\Delta \lambda_{\rm inst, EIS}' = 62.7$\,m\mbox{\AA} for the $2\arcsec$ slit at the averaged 30 pixels, which is about 9\% smaller than the original instrumental width $\Delta \lambda_{\rm inst, EIS} = 69.7$\,m\mbox{\AA} and beyond the 3\,m\mbox{\AA} uncertainty. The difference between the two EIS instrumental widths $(\Delta \lambda_{\rm inst, EIS}^2 - \Delta \lambda_{\rm inst, EIS}'^2 )^{1/2} = 30.4\,$m\mbox{\AA} is equivalent to an effective velocity of $28.3\,\mathrm{km \, s^{-1}}$, which might be caused by Fe \textsc{xii} nonthermal motions not properly removed in \citet{EISNote7}. \citet{DelZanna2019} also reported the possible overestimation of EIS instrumental width. 

\begin{figure*}
    \centering
    \includegraphics[width=\linewidth]{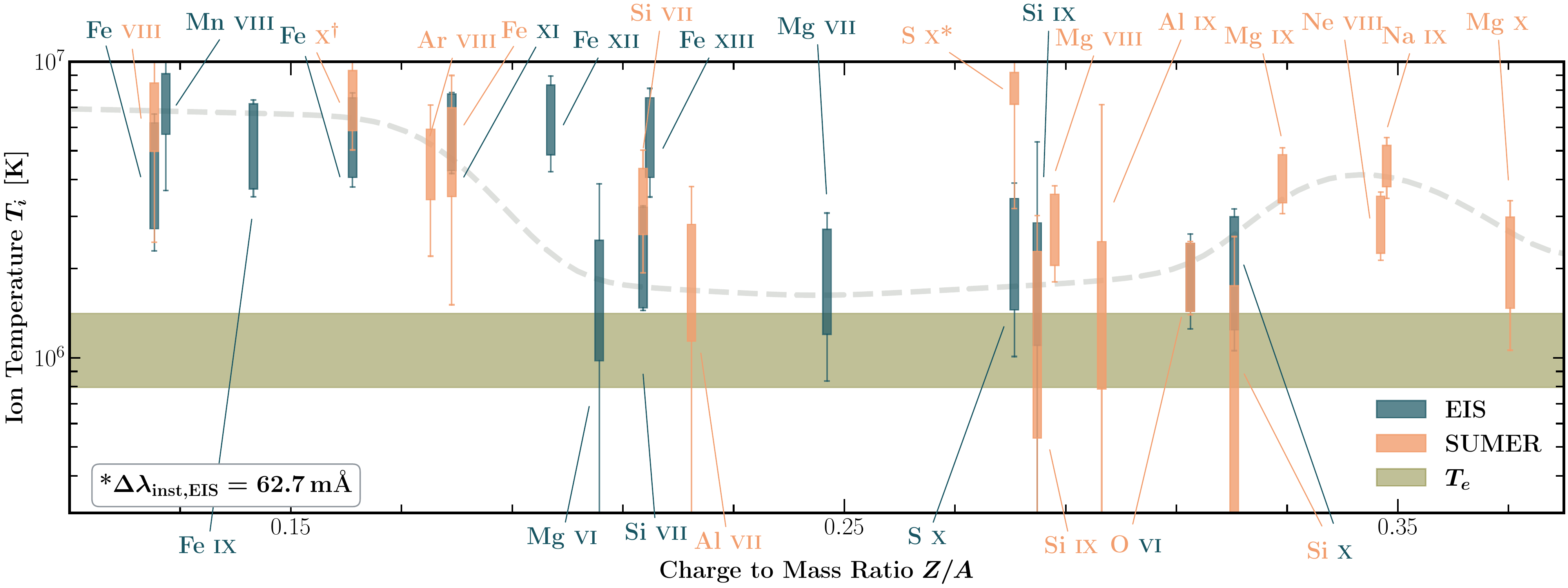}
    \caption{Same as Figure~\ref{fig:temp_diag} but using the cross-calibrated EIS instrumental width $\Delta \lambda_{\rm inst,EIS}' = 62.7$\,m\mbox{\AA}. The dashed gray line is arbitrarily drawn to illustrate the U-shaped dependence. Link to the \texttt{Jupyter} notebook creating this figure: \href{https://github.com/yjzhu-solar/EIS_SUMER_PCH_Ti/blob/main/ipynb/paper/temp_diag_v2_cross.ipynb}{\faGithub}.}
    \label{fig:temp_diag_cross}
\end{figure*}

We repeated ion temperature diagnostics with the new EIS instrumental width $\Delta \lambda_{\rm eff, EIS}' = 62.7$\,m\mbox{\AA} in Figure~\ref{fig:temp_diag_cross}. The lowest $v_{\rm eff}$ is from the Si \textsc{x} 624.694\,\mbox{\AA} line observed by SUMER, which has a more considerable fitting uncertainty due to the blended Mg \textsc{x} 624.941\,\mbox{\AA} line. The EIS ion temperatures $T_i$ at low $Z/A$ increase significantly and overlap the $T_i$ ranges of the same ion observed by SUMER (e.g., Fe \textsc{viii} and Fe \textsc{xi}). The increase in the EIS $T_i$ also makes the U-shaped dependence on $Z/A$ more prominent. The $T_i$ of the Fe \textsc{xii} and Fe \textsc{xiii} observed by EIS are much higher than the $T_i$ of Mg \textsc{vi}, Si \textsc{vii}, and Al \textsc{vii} with similar $Z/A$ between 0.20 and 0.23. This might be due to the macroscopic Doppler broadening of ions with higher $T_{\rm max}$ found in AWSoM-R simulations (see Sections~\ref{subsec:simulations} and \ref{subsec:valid}). 

\begin{figure*}
    \centering
    \includegraphics[width=\linewidth]{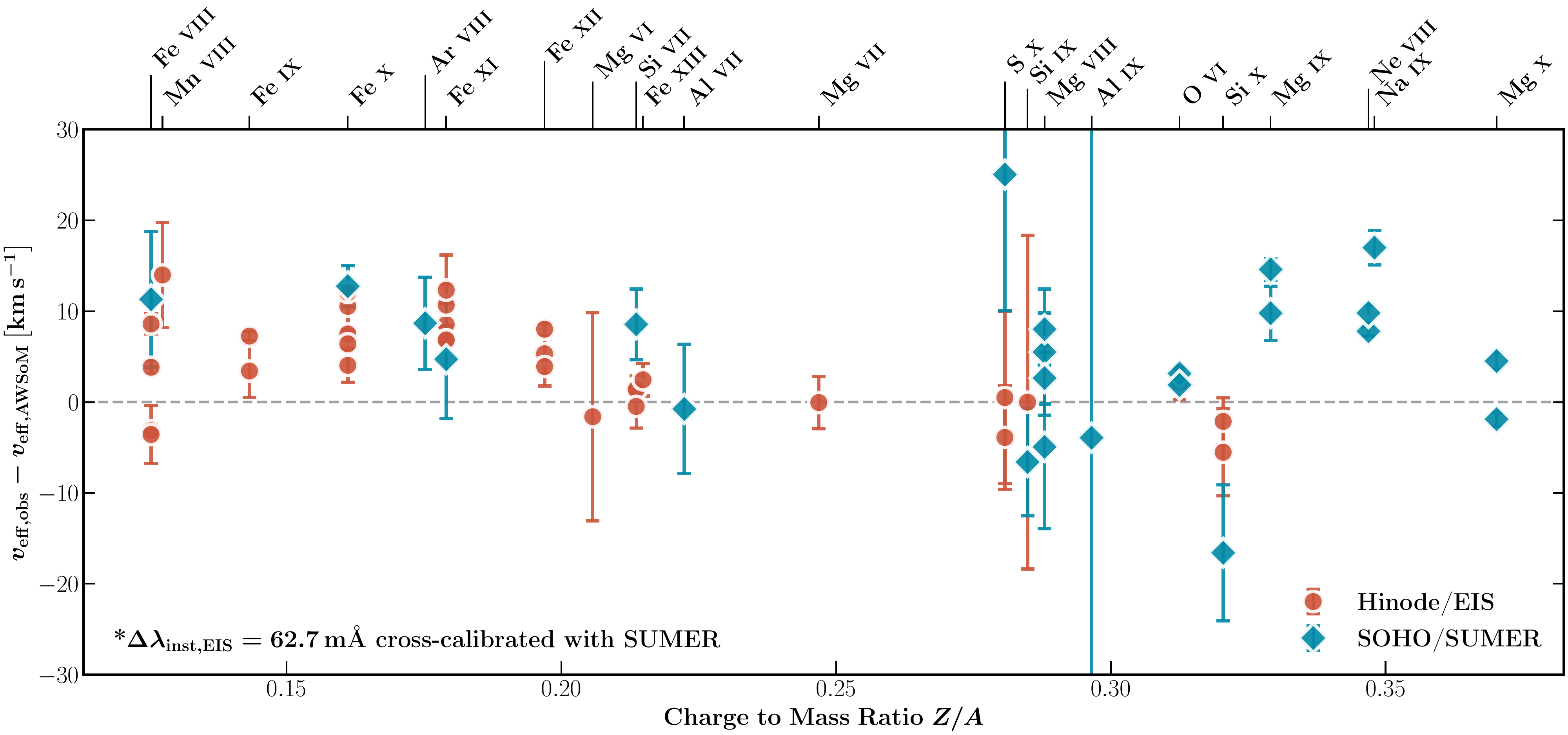}
    \caption{Same as Figure~\ref{fig:awsom_obs} but using EIS instrumental broadening $\Delta \lambda_{\rm inst,EIS}' = $62.7\,m\mbox{\AA} cross-calibrated with SUMER. Link to the \texttt{Jupyter} notebook creating this figure: \href{https://github.com/yjzhu-solar/EIS_SUMER_PCH_Ti/blob/main/ipynb/sim_obs_comp/sim_obs_linewidth.ipynb}{\faGithub}.}
    \label{fig:awsom_obs_cross}
\end{figure*}

Figure~\ref{fig:awsom_obs_cross} compares the observed line widths and the synthetic line widths using the cross-calibrated instrumental width of EIS. The $v_{\rm eff}$ of the EIS lines increase by $\sim 5$\textendash$10\,\mathrm{km \, s^{-1}}$. The EIS lines with $0.12\leq Z/A \leq 0.19$ are about 0\textendash$15\,\mathrm{km \, s^{-1}}$ wider than the synthetic lines, which is consistent with the SUMER observation. The EIS lines with $0.2\leq Z/A\leq 0.32$ show similar line widths compared to the synthetic ones (within $\pm 5\,\mathrm{km \, s^{-1}}$). 

\subsection{Validation of the Technique } \label{subsec:valid}

\begin{figure*}
    \centering
    \includegraphics[width=\linewidth]{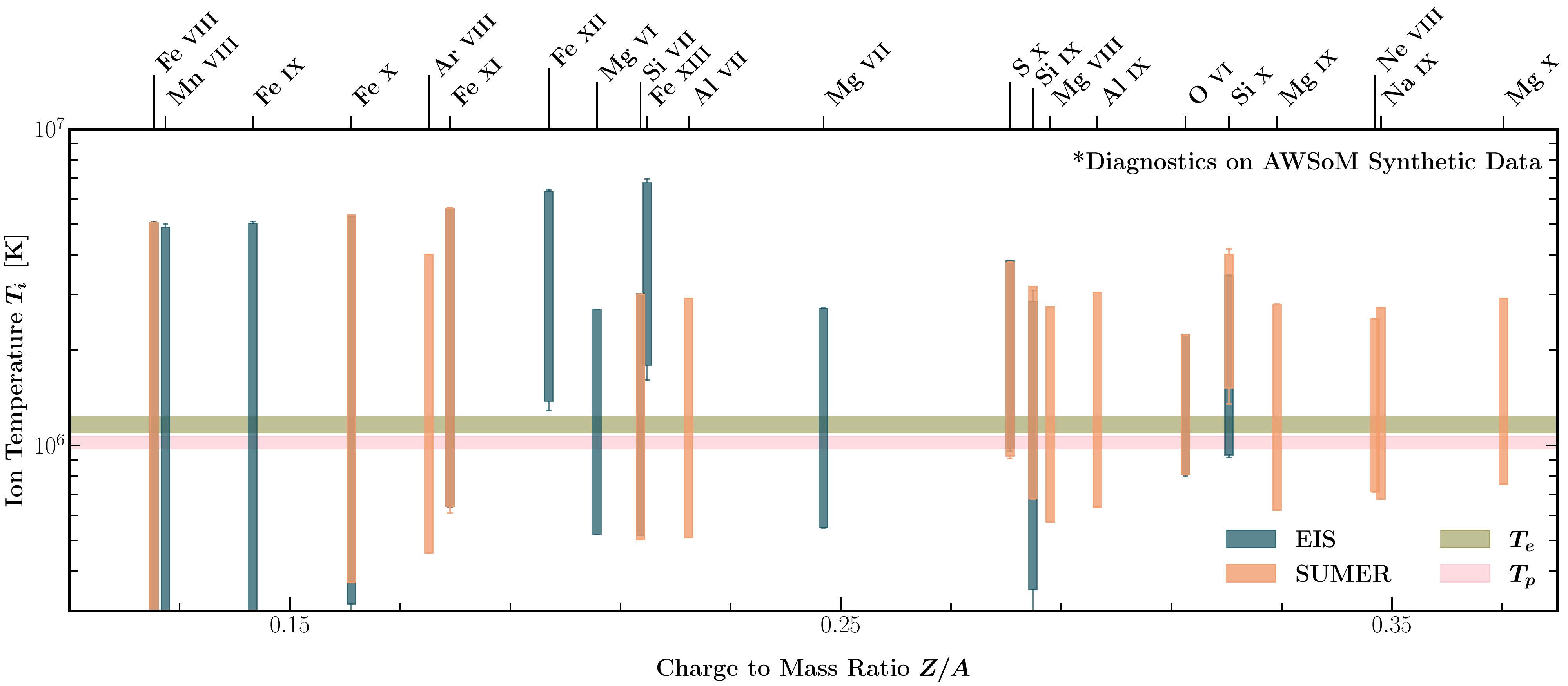}
    \caption{Same as Figure~\ref{fig:temp_diag}, but using line widths from AWSoM-R simulations. The colored horizontal area represents the range of average electron temperature $T_e$ and proton temperature $T_p$ along the LOS weighted by either Fe \textsc{viii} 186\,\mbox{\AA} emissivity or Fe \textsc{xii} 192\,\mbox{\AA} emissivity. Link to the \texttt{Jupyter} notebook creating this figure: \href{https://github.com/yjzhu-solar/EIS_SUMER_PCH_Ti/blob/main/ipynb/paper/temp_diag_pseudo.ipynb}{\faGithub}.}
    \label{fig:temp_diag_awsom}
\end{figure*}

To validate the $T_i$ diagnostics technique, we performed the same diagnostics on the AWSoM-R synthetic line widths. The SPECTRUM module uses the proton temperature $T_p$ to calculate the thermal widths of all spectral lines. Hence, the measured $T_i$ should show no dependence on $Z/A$. Figure~\ref{fig:temp_diag_awsom} shows the diagnostic results along with the weighted average of the electron temperature $\overline{T}_{\!e}$ and the proton temperature $\overline{T}_{\! p}$ along the LOS. We used the emissivity $\epsilon_{ij}$ of Fe \textsc{viii} 186\,\mbox{\AA} and Fe \textsc{xii} 192\,\mbox{\AA} line as the weights:
\begin{equation}
    \overline{T} = \frac{\int \epsilon_{ij} (x) T(x)\mathrm{d}x}{\int \epsilon_{ij} (x)\mathrm{d}x} 
\end{equation}
to determine the interval of the weighted average $\overline{T}_{\!e}$ and $\overline{T}_{\!p}$. 

As expected, the measured $T_i$ intervals $[T_{i,\mathrm{min}},T_{i,\mathrm{max}}]$ do not show U-shaped relations with $Z/A$ and are consistent with both $\overline{T}_e$ and $\overline{T}_p$, because there is no preferential heating in the simulation. The only exceptions are the Fe \textsc{xii} and Fe \textsc{xiii} due to the bulk motions along the LOS. Hence, we validated that the $T_i$ diagnostic technique can be used to search for the preferential heating of heavy ions, especially for the cooler ions observed by SUMER and EIS (e.g., Fe \textsc{viii} and Ne \textsc{viii}).

 
\subsection{Non-Gaussian Profiles} \label{subsec:nongauss}

The line-fitting and $T_i$ diagnostics assume that the observed line profiles are close to Gaussian. However, we found that the brightest lines observed by SUMER ($I_{\rm tot} > 1\,\mathrm{erg \, s^{-1} \, cm^{-2}}$\,\mbox{\AA}$^{-1}\,\mathrm{sr^{-1}}$) show non-Gaussian wings at the coronal hole boundary. These lines include O \textsc{vi}, Ne \textsc{viii}, Mg \textsc{ix}, and Mg \textsc{x}. The non-Gaussian wings in the plumes and interplume regions were reported in early SUMER observations \citep[e.g.,][]{Hassler1997,Wilhelm1999}. \citet{Wilhelm1999} found non-Gaussian wings of the Ne \textsc{viii} 780\,\mbox{\AA} line in the coronal hole; on the contrary, the Ne \textsc{viii} profiles in closed magnetic field regions do not show broad non-Gaussian wings. They also suggested that the non-Gaussian wings are not instrumental effects because the brightest C \textsc{iii} stray-light line does not show the non-Gaussian wings. Similarly, we did not find non-Gaussian wings in the brightest N \textsc{iv} 765\,\mbox{\AA} stray-light line in this data set either.

To examine the influence of the non-Gaussian wings on the single-Gaussian fitting, we first fitted the non-Gaussian profile by the Voigt function or a secondary Gaussian component. We found that the double-Gaussian function better reproduces the non-Gaussian wings, consistent with negligible pressure broadening in the coronal holes to produce Voigt profiles. 

\begin{figure*}
\gridline{\fig{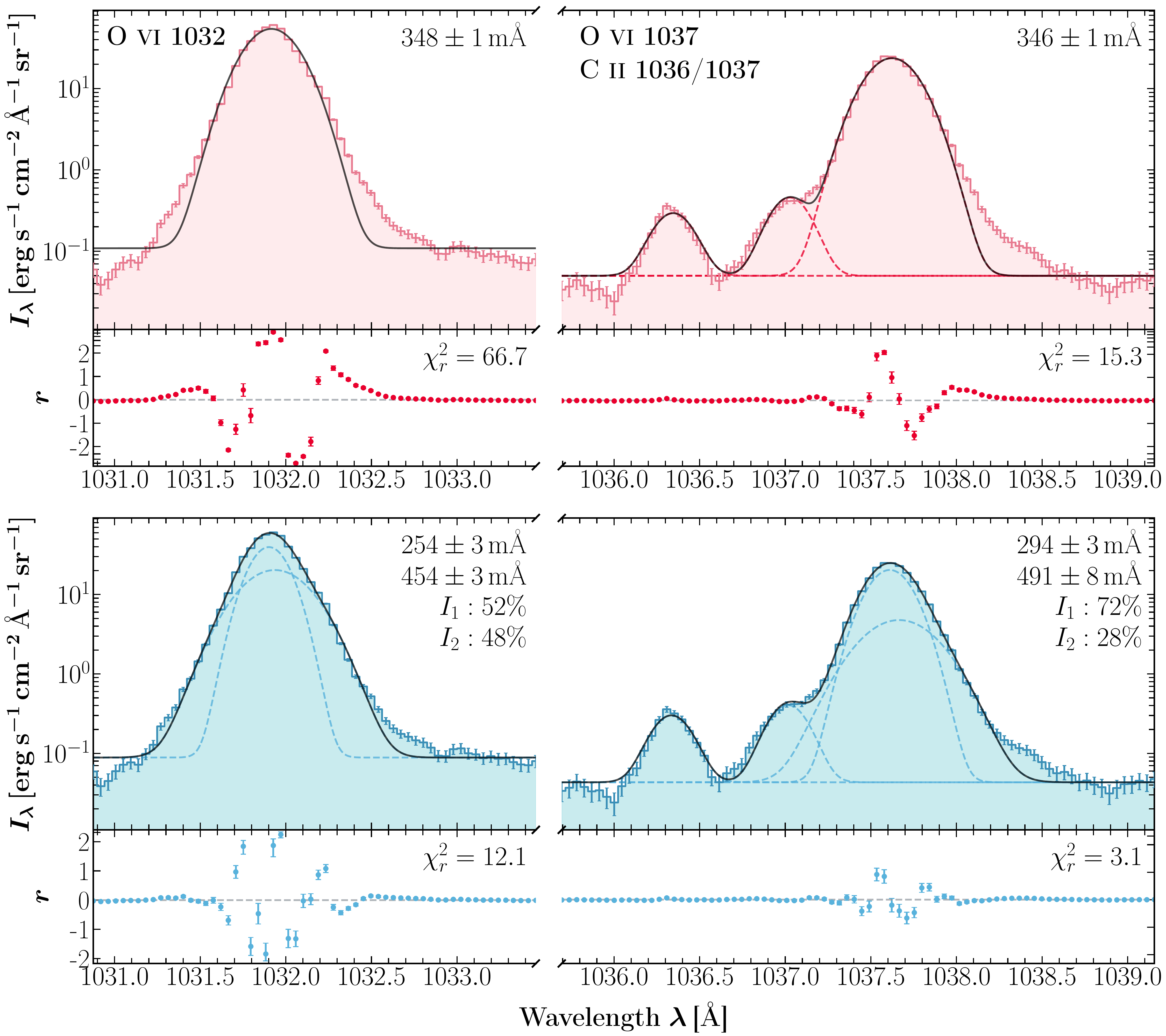}{0.49\textwidth}{(a)}
          \fig{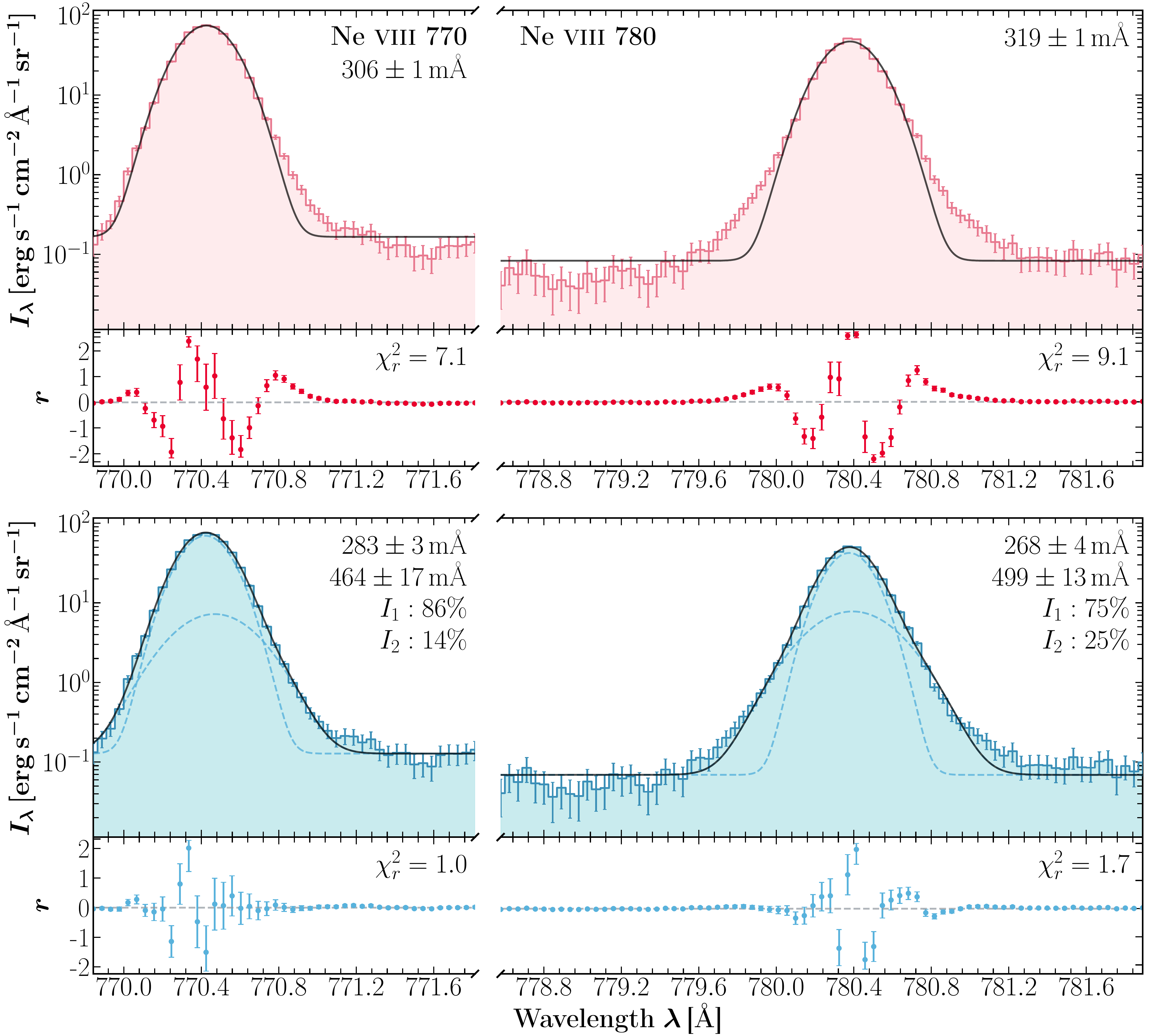}{0.49\textwidth}{(b)}}
\gridline{\fig{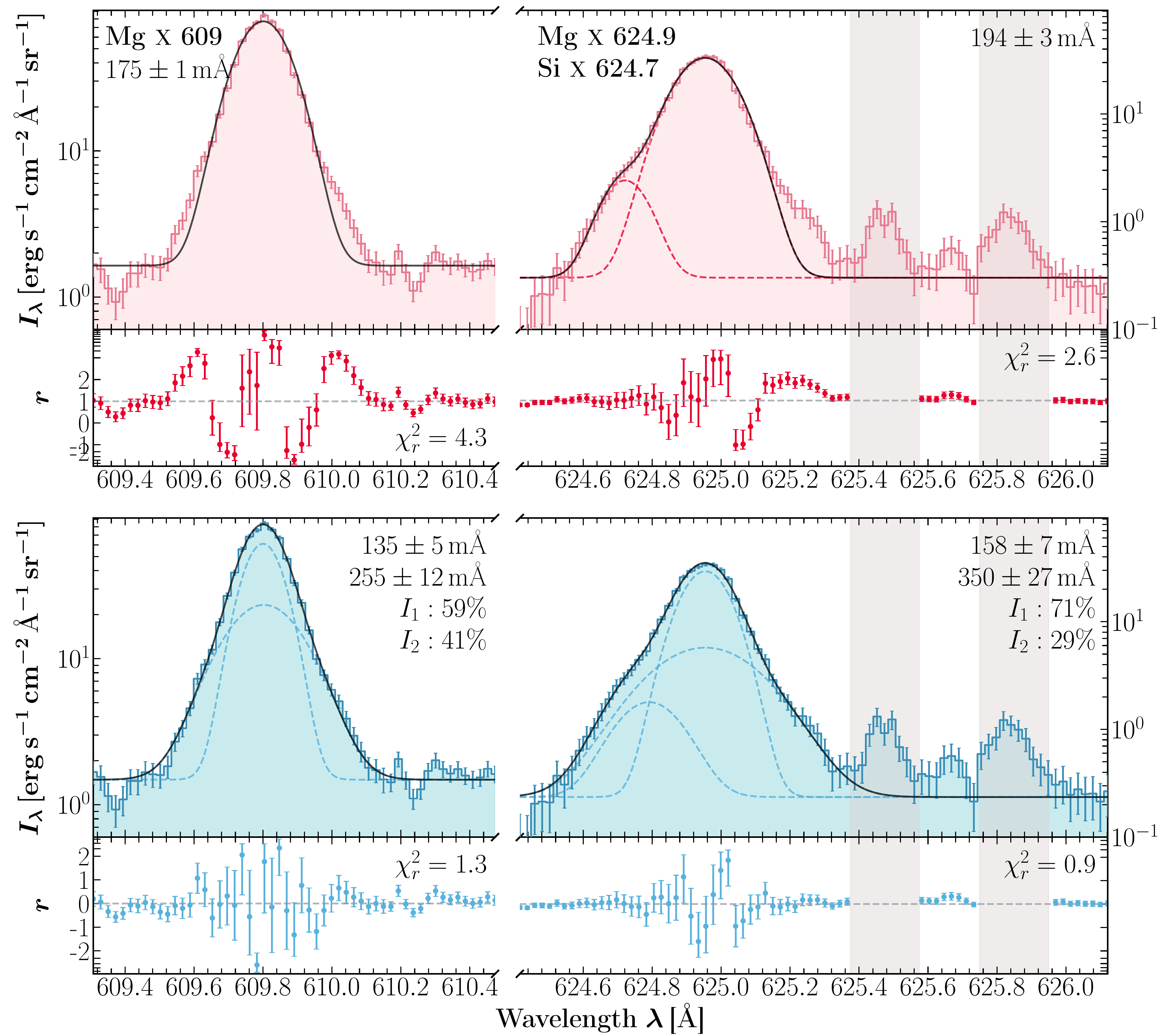}{0.49\textwidth}{(c)}}
    \caption{Single-Gaussian and double-Gaussian fitting to the \textbf{(a)} O \textsc{vi} 1032 and 1037\,\mbox{\AA}, \textbf{(b)} Ne \textsc{viii} 770 and 780\,\mbox{\AA}, and \textbf{(c)} Mg \textsc{x} 609 and 624\,\mbox{\AA} lines, including the adjacent C \textsc{ii} 1036 and 1037\,\mbox{\AA} and Si \textsc{x} 624\,\mbox{\AA} lines. The black solid curve shows the fitting result. The dashed curves show each Gaussian component. The single-Gaussian widths, the double-Gaussian widths, and relative intensities are also shown. The vertical gray areas in panel c indicate the masked pixels during the fitting. Note that we only fixed the line centroids in the double-Gaussian fitting of the Mg \textsc{x} 624\,\mbox{\AA} line. Link to the \texttt{Jupyter} notebook creating this figure: \href{https://github.com/yjzhu-solar/EIS_SUMER_PCH_Ti/blob/main/ipynb/paper/non_gauss_profile.ipynb}{\faGithub}.}
    \label{fig:double_gauss}
\end{figure*}

We compared the single-Gaussian and double-Gaussian fitting results of O \textsc{vi} 1032/1037\,\mbox{\AA}, Ne \textsc{viii} 770/780\,\mbox{\AA}, and Mg \textsc{x} 609/624\,\mbox{\AA} in Figure~\ref{fig:double_gauss}. The secondary Gaussian component improves the fitting up to $\sim \pm 200\,\mathrm{km \, s^{-1}}$, but still leaves some residuals in the red wings of the O \textsc{vi} and Ne \textsc{viii} lines.

After removing instrumental broadening, the width of the narrower main component in the double-Gaussian fitting is 10\%\textendash40\% less than the single-Gaussian width, equivalent to a reduction of 20\%\textendash60\% in $T_{i,\mathrm{max}}$. On the other hand, the broader secondary Gaussian component has a width about twice that of the primary component, corresponding to an effective temperature 2\textendash3 times that of the single-Gaussian profile. Since none of these brightest lines is used to estimate the maximum nonthermal velocity $\xi_{\rm max}$, most $T_{i,\mathrm{min}}$ measurements will not be affected. 

The intensity ratios between the two Gaussian components suggested that the flows along the LOS might not be the only cause of the non-Gaussian wings. If unresolved flows were present, the intensity ratio should be the same for both lines of the same ion. On the contrary, these ratios do not agree, let alone with the ratios from other ions. For example, in O \textsc{vi} 1032\,\mbox{\AA} and Mg \textsc{x} 609\,\mbox{\AA}, the secondary component contributes approximately half the intensity of the line profile, while the contribution of the secondary component in O \textsc{vi} 1037\,\mbox{\AA} and Mg \textsc{x} 624\,\mbox{\AA} is much less than 50\%. 

The inconsistency of the intensity ratios between the two Gaussian components raises the question of whether the double-Gaussian fitting is a good approximation to fit the high-energy tails. The $\kappa$ distribution, first introduced empirically by \citet{Vasyliunas1968} and \citet{Olbert1968}, is helpful in fitting the suprathermal tails of plasma particles \citep{Lazar2016}.
Inspired by \citet{Jeffrey2018} who fitted EIS spectra in the coronal hole assuming a $\kappa$ distribution, we also attempted fitting these brightest SUMER lines with the $\kappa$ distribution using the formula in \citet{Dudik2021}:
\begin{equation}
    I_\kappa (\lambda) = I_0 \left[1 + \frac{(\lambda - \lambda_0)^2}{2(\kappa - 3/2)w_\kappa^2} \right]^{-\kappa} 
\end{equation}
where $I_0$ is the peak intensity and $w_\kappa$ represents the characteristic width. The $w_\kappa$ is related to the FWHM $\Delta \lambda_\kappa$ of $\kappa$ distribution by \citep[see][]{Dudik2017}
\begin{equation}
    w_\kappa = \frac{1}{8}\frac{\Delta \lambda_\kappa^2}{(\kappa - 3/2)(2^{1/\kappa} - 1)}
\end{equation}

\begin{figure}
    \centering
    \includegraphics[width=\linewidth]{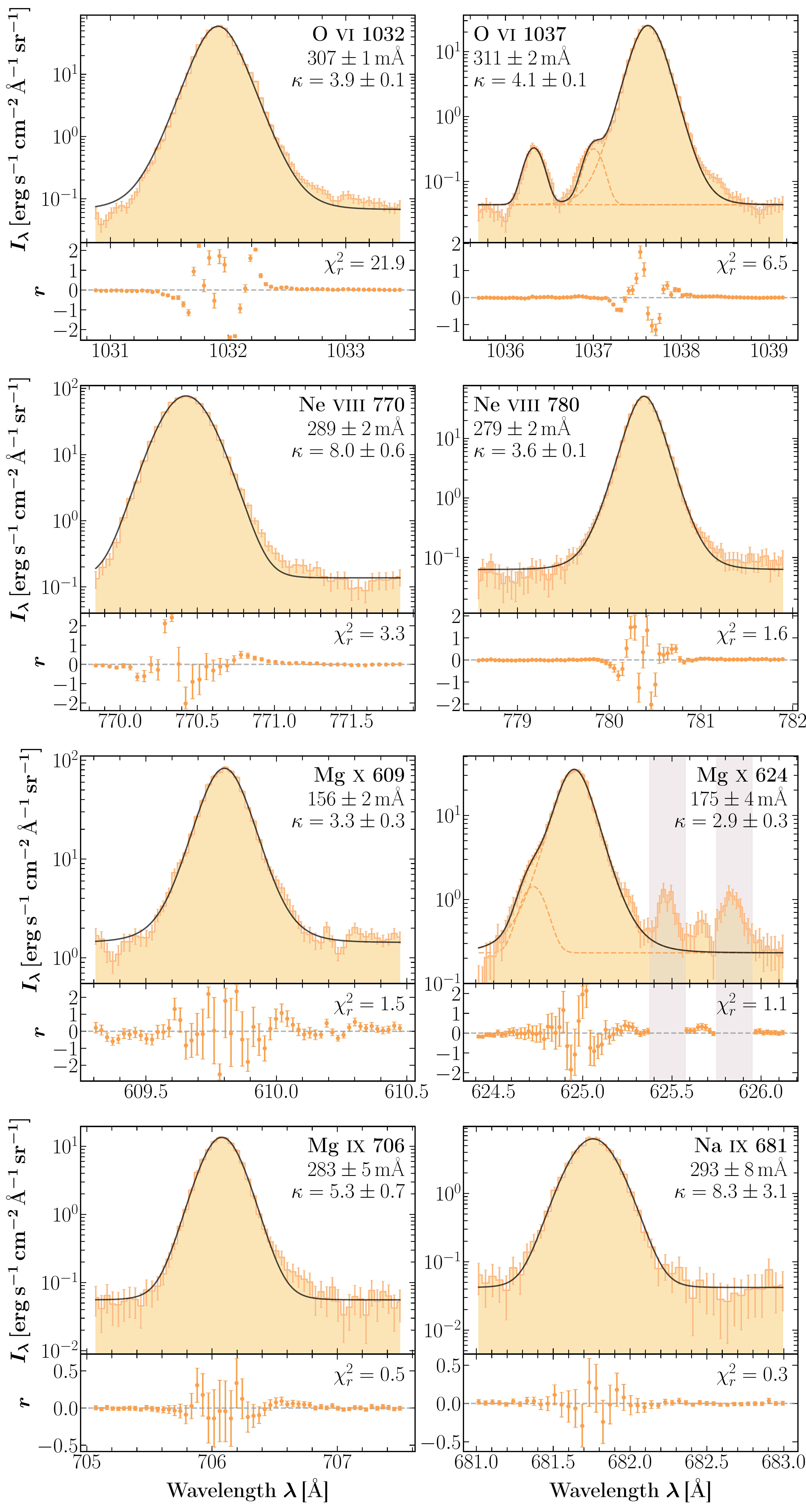}
    \caption{$\kappa$ fitting of the brightest coronal lines observed by SUMER. The fitted $\kappa$ and FWHM of the $\kappa$ profile $\Delta \lambda_\kappa$ are shown as well. The weak blended or stray-light lines are still fitted with a Gaussian distribution. Link to the \texttt{Jupyter} notebook creating this figure: \href{https://github.com/yjzhu-solar/EIS_SUMER_PCH_Ti/blob/main/ipynb/paper/non_gauss_profile.ipynb}{\faGithub}.}
    \label{fig:kappa_fit}
\end{figure}

Figure~\ref{fig:kappa_fit} shows the SUMER line profiles fitted with the $\kappa$ distribution. The $\kappa$ distribution well fitted the non-Gaussian wings. The six brightest resonant lines of O \textsc{vi}, Ne \textsc{viii}, and Mg \textsc{x} show similar fitted $\kappa \sim 3$\textendash4, except for the Ne \textsc{viii} 770\,\mbox{\AA} with $\kappa \sim 8$. The fainter Mg \textsc{ix} 706\,\mbox{\AA} and Na \textsc{ix} 681\,\mbox{\AA} show a larger $\kappa\sim 5$ and $\kappa\sim 8$, respectively. These $\kappa$ values are slightly greater than the $\kappa \approx 1.9$\textendash2.5 obtained from EIS observations of the southern polar coronal hole reported by \citet{Jeffrey2018}.   

Since the SUMER software is designed to remove the instrumental broadening from Gaussian profiles rather than from $\kappa$ profiles, we roughly estimated the influence of $\kappa$ fitting on $T_{i,\mathrm{max}}$ using the line width before the instrumental broadening correction \citep[][]{Dudik2017}:
\begin{equation}
    \frac{T^{G}_{i,\mathrm{max}}}{T^\kappa_{i,\mathrm{max}}} \sim \frac{(\kappa - 3/2)(2^{1/\kappa} - 1)}{\ln 2}\frac{\Delta \lambda^2_G}{\Delta \lambda^2_\kappa}
\end{equation}
where $T^{G}_{i,\mathrm{max}}$ is the maximum ion temperature estimated from the Gaussian FWHM $\Delta \lambda_G$, and $T^{\kappa}_{i,\mathrm{max}}$ is the maximum ion temperature estimated from the $\kappa$ FWHM $\Delta \lambda_\kappa$. The $T^{G}_{i,\mathrm{max}}/T^{\kappa}_{i,\mathrm{max}}$ ratios of unblended O \textsc{vi} 1032\,\mbox{\AA}, Ne \textsc{viii} 770 and 780\,\mbox{\AA}, and Mg \textsc{x} 609\,\mbox{\AA} are between 0.75 and 0.95, which means $T_{i,\mathrm{max}}$ might increase by 10\%\textendash20\% if we take the high-energy tails into account.

We agree with \citet{Jeffrey2018} that nonequilibrium ion populations, non-Gaussian turbulence, or both might cause the non-Gaussian wings in the coronal hole. We did not make further investigations into the formation mechanism of the high-energy tails, which is out of the scope of this paper. 

\subsection{Preferentially Heated Ions}\label{subsec:diss_heating}
\begin{figure*}
    \centering
    \includegraphics[width=\textwidth]{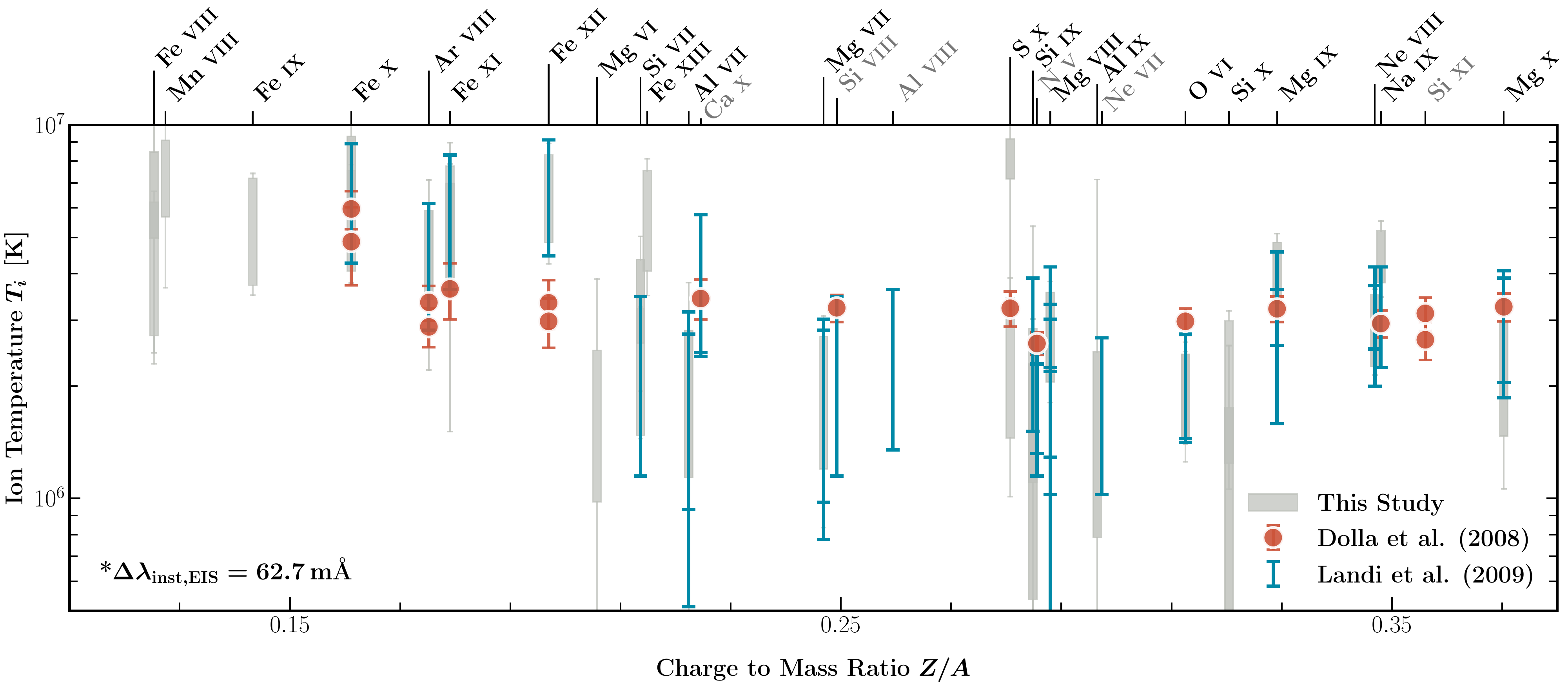}
    \caption{Comparison between the ion temperature $T_i$ in the polar coronal hole measured by this study (gray box plot), \citet[][red error bars]{Dolla2008}, and \citet[][blue error bars]{Landi2009}. Gray ion names represent ions that are not used in this study. Link to the \texttt{Jupyter} notebook creating this figure: \href{https://github.com/yjzhu-solar/EIS_SUMER_PCH_Ti/blob/main/ipynb/paper/temp_diag_v2_cross.ipynb}{\faGithub}.}
    \label{fig:temp_landi_dolla}
\end{figure*}

Figure~\ref{fig:temp_landi_dolla} compares our ion temperature $T_i$ measurements at $\sim 1.03\,R_\odot$ at the coronal hole boundary with two previous studies: \citet[][1.06\,$R_\odot$]{Landi2009} and \citet[][1.05\,$R_\odot$]{Dolla2008}, which used SUMER observations at the center of polar coronal holes to measure $T_i$. \citet{Landi2009} also used the diagnostic method, while \citet{Dolla2008} separated the thermal and nonthermal widths by assuming the thermal width of Mg \textsc{x} 624\,\mbox{\AA} is constant at different altitudes and the variation of the nonthermal width is caused by undamped Alfv\'en waves.

We confirmed the U-shape dependence of $T_i$ on $Z/A$ in the polar coronal hole and the preferential heating of the ions with $Z/A<0.2$ or $Z/A>0.33$. Furthermore, we extended this result to $Z/A < 0.15$. The heating of the $Z/A > 0.33$ ions is inconsistent with the traditional cascade model of ion-cyclotron waves and implies additional resonant wave power at high frequencies \citep[large $Z/A$,][]{Landi2009}. We note that the high-$Z/A$ ions are only observed by SUMER, which makes the preferential heating at $Z/A \geq 0.33$ less robust than the heating at $Z/A \leq 0.19$ confirmed by both SUMER and EIS.

Although EIS observes most of the lower-$Z/A$ lines used in this study, using the cross-calibrated EIS instrumental width, our results show remarkable consistency with the SUMER-based $T_i$ values reported by \citet{Landi2009}. On the other hand, some of the $T_i$ measured by \citet{Dolla2008} do not fall within the $T_i$ interval found by this study, for example, Ar \textsc{viii}, Fe \textsc{x}, Fe \textsc{xii}, and Na \textsc{ix} are found to be hotter in this study. We note that both \citet{Landi2009} and this study used polar coronal hole observations during the solar minimum. In contrast, \citet{Dolla2008} analyzed the SUMER observations taken during the solar maximum, which might cause the differences in the measured $T_i$, as the ion charge state in the fast solar wind is found to vary from solar maximum to minimum \citep[e.g.,][]{Lepri2013}. 

\citet{Hahn2010} analyzed the same EIS data set and applied the same $T_i$ diagnostics from 1.04 to 1.14\,$R_\odot$. They found similar preferential heating of ions with $Z/A < 0.2$. However, the heating of $Z/A > 0.33$ ions (e.g., Mg \textsc{ix} and Ne \textsc{viii}) was missing in their study because only SUMER can observe the spectral lines of these ions. The EIS $T_i$ intervals measured using the cross-calibrated instrumental width are slightly lower than the ion temperature $\log T_{i,\mathrm{max}} > 7.0$ reported by \citet{Hahn2010}. This is because \citet{Hahn2010} measured $T_i$ at higher altitudes ($\geq 1.04\,R_\odot$) and used an old EIS instrumental width of 61\,m\mbox{\AA} (SW detector) or 62\,m\mbox{\AA} (LW detector).  

\section{Conclusion} \label{sec:conclusion}
The heavy-ion temperatures $T_i$ provide essential information about the heating mechanism of the million-degree corona. In this study, we estimated possible $T_i$ intervals $\left[T_{i,\rm{min}},T_{i,\rm{max}} \right]$ at the polar coronal hole boundary simultaneously observed by Hinode/EIS and SOHO/SUMER. We studied the dependence of $T_i$ on the heavy-ion charge-to-mass ratios $(Z/A)$ between 0.125 and 0.37 and compared $T_i$ with the local electron temperature $T_e$. We further validated our $T_i$ diagnostic results using the line profiles synthesized from AWSoM-R. 

We found that heavy ions with $0.12 < Z/A < 0.2$ and $0.33 < Z/A < 0.35$ are preferentially heated at the base of the coronal hole boundary. $T_{i,\rm{min}}$ of the preferentially heated ions are greater than $T_e$ by a factor of 1.5 - 3. The $T_i$ intervals show a nonmonotonic, U-shaped dependence on $Z/A$ of heavy ions, which is inconsistent with the traditional cascade models of the ion-cyclotron resonance \citep{Landi2009}.

We found the EIS instrumental width is one of the most significant contributors to the uncertainty of the $T_i$ measurement. The spectral lines observed by SUMER are $\sim 30\%$ broader than the lines originating from the same ion observed by EIS when we used the recommended EIS instrumental width $\Delta \lambda_{\rm inst,EIS} = 69.7\,\mathrm{m} \mbox{\AA}$. We derived a narrower EIS instrumental width $\Delta \lambda_{\rm inst,EIS} = 62.7\,\mathrm{m} \mbox{\AA}$ by comparing the widths of O \textsc{vi} 184.1\,\mbox{\AA} and O \textsc{vi} 1032/1037\,\mbox{\AA} lines. The new instrumental widths provide more consistent $\left[T_{i,\rm{min}},T_{i,\rm{max}} \right]$ measurements between EIS and SUMER. 

The AWSoM-R simulation validated the preferential heating of the heavy ions and $T_i$ diagnostic techniques. The synthetic lines of the preferentially heated ions are narrower than the observed ones, probably because the ion-cyclotron resonance and the heavy-ion temperature are not modeled in AWSoM-R. The AWSoM-R simulation also suggests that the line profiles from hot ions, such as Fe \textsc{xii} and Fe \textsc{xiii}, might be affected by bulk velocity along the LOS. 

We confirmed some of the brightest spectral lines observed by SUMER show enhanced, non-Gaussian wings in the coronal hole, including O \textsc{vi}, Ne \textsc{viii}, and Mg \textsc{x}. Compared to the double-Gaussian function, the $\kappa$ distribution fits the lines profiles better with a $\kappa \sim 3$\textendash4. If the high-energy tails are related to the thermal velocity of heavy ions, the estimated $T_{i,\rm{max}}$ might increase by another 10\%\textendash20\%. 

Our study reveals the complicated dependence of the $T_i$ on the $Z/A$, which is vital to assessing coronal heating theories. We encourage future studies of the coronal $T_i$ using EUV spectral lines observed by EIS, the Spectral Imaging of the Coronal Environment \citep[SPICE;][]{SpiceConsortium2020} instrument with the corrected point spread function (PSF), and the upcoming Solar-C (EUVST) mission \citep{Shimizu2019}, as well as the visible and infrared forbidden lines observed during the eclipse \citep[e.g.,][]{Ding2017} and the NSF's Daniel K. Inouye Solar Telescope \citep[DKIST;][]{Rimmele2020}. 

\section*{}
The authors thank the anonymous referee for constructive suggestions and comments. Funding for the DKIST Ambassadors program is provided by the National Solar Observatory, a facility of the National Science Foundation, operated under Cooperative Support Agreement number AST-1400405. The work of EL was supported by NASA grants 80NSSC20K0185, 80NSSC21K0579, 80NSSC18K1553, 80NSSC18K0647, 80NSSC22K0750, and 80NSSC22K1015, and NSF grant AGS-2229138. SOHO is a project of international cooperation between ESA and NASA. The SUMER project is financially supported by DLR, CNES, NASA, and the ESA PRODEX Program (Swiss contribution). Hinode is a Japanese mission developed and launched by ISAS/JAXA, collaborating with NAOJ as a domestic partner, NASA and UKSA as international partners. Scientific operation of the Hinode mission is conducted by the Hinode science team organized at ISAS/JAXA. This team mainly consists of scientists from institutes in the partner countries. Support for the post-launch operation is provided by JAXA and NAOJ (Japan), UKSA (U.K.), NASA, ESA, and NSC (Norway). This work utilizes data obtained by the Global Oscillation Network Group (GONG) program, managed by the National Solar Observatory, which is operated by AURA, Inc.  under a cooperative agreement with the National Science Foundation.  The data were acquired by instruments operated by the Big Bear Solar Observatory, High Altitude Observatory, Learmonth Solar Observatory, Udaipur Solar Observatory, Instituto de Astrofisica de Canarias, and Cerro Tololo Interamerican Observatory. CHIANTI is a collaborative project involving George Mason University, the University of Michigan (USA) and the University of Cambridge (UK). The authors thank Dr. Michael Hahn for helpful discussions on the EIS instrumental broadening. The authors also acknowledge high-performance computing support from Pleiades, operated by NASA's Advanced Supercomputing Division. 

%

\facilities{Hinode(EIS), SOHO(SUMER and EIT)}


\software{Numpy \citep{oliphant2006guide,van2011numpy}, Scipy \citep{2020SciPy-NMeth}, Astropy \citep{2013A&A...558A..33A,2018AJ....156..123A}, Sunpy \citep{sunpy_community2020}, Matplotlib \citep{Hunter2007}, corner.py \citep{corner}, CHIANTI \citep{Dere1997,DelZanna2021}, SolarSoft \citep{Freeland2012}, num2tex\footnote{\url{https://github.com/AndrewChap/num2tex}}, cmcrameri \citep{Crameri2021}, brokenaxes\footnote{\url{https://github.com/bendichter/brokenaxes}}. The \texttt{Jupyter} notebooks and \texttt{IDL} scripts for data reduction, analysis, and visualization are available on Zenodo \citep{CodeRepoZhu2023} and GitHub.}



\appendix
\restartappendixnumbering
\section{Data Calibration and Coalignment}\label{append:calib_coalign}

We retrieved the SUMER data from the original telemetry through the SUMER Image Database. Then we applied the standard data corrections and calibrations described in the SUMER Data Cookbook, including the decompression, reversion, dead-time correction, flat-field correction, local-gain correction, and geometric distortion correction. We determined the illuminated portion of the $1024\times360$ SUMER detector by manually checking the intensity distribution along the $y$ direction in the four spectral windows. Then we resized the images to $1024\times300$ using the IDL \texttt{congrid} function. We calculated the uncertainty of the SUMER intensity in each pixel, assuming that the uncertainty is dominated by the photon shot noise following Poisson statistics \citep[][]{Peter1999}, namely, 
\begin{equation}
    \sigma_P = \sqrt{P}
\end{equation}
where $P$ is the total photon counts per pixel and $\sigma_P$ is the corresponding uncertainty. The uncertainty of the radiometrically calibrated intensity $I$ is given by \citet{EISNote1}:
\begin{equation}
    \frac{\sigma_I}{I} = \frac{\sigma_P}{P}
\end{equation} 
Since SUMER does not provide an absolute wavelength calibration, we calibrated the wavelength across the detector by performing linear regression between line centroids (in detector pixels) and the wavelengths provided by the CHIANTI atomic database \citep{Dere1997,DelZanna2021}. The difference between the calibrated pixel sizes and the pixel sizes given by the grating dispersion relation in the SUMER software is less than 0.5\%. No absolute wavelength calibration is performed, as we only used the line width and intensity in the following diagnostics. Finally, we applied the latest (Epoch 9) SUMER radiometric calibration to the individual spectral line before the fitting.

We obtained the EIS level-0 FITS files from the Hinode Science Data Center Europe Archive. We first calibrated the EIS data set to level 1 using the IDL routine \texttt{eis\_prep} \citep{EISNote1}. The level 1 data set contains the intensity after the original laboratory radiometric calibration. Then we applied additional corrections to the level-1 data set, including the Y-offset of the CCD \citep{EISNote3} and the tilt of the slit \citep{EISNote4}. Since $T_i$ diagnostics only relies on line widths, additional radiometric corrections \citep[i.e.,][]{DelZanna2013,Warren2014} were only implemented on the intensity of the spectral lines used for electron density and temperature diagnostics (see Section~\ref{subsec:electemp} for details) after fitting. We averaged all seven EIS rasters to increase the S/N.  

For context, we checked the global solar corona images taken by SOHO/EIT. SUMER, EIS, and EIT were coaligned manually in two steps: (1) We coaligned EIT and EIS by comparing the on-disk features in the EIT 195\,\mbox{\AA} image and the EIS SW spectrum filtered by the effective area of the EIT 195\,\mbox{\AA} quadrant. (2) We empirically coaligned SUMER and EIS by comparing the slope of the O \textsc{vi} 184.1\,\mbox{\AA} (EIS) and 1031.9\,\mbox{\AA} (SUMER) intensities along the slit. No coalignment in the east\textendash west direction was performed between the EIS and SUMER data. The uncertainty of the coalignment between EIT and EIS is less than $5\arcsec$ (EIT pixel size) since the features on the disk are well matched. However, since the off-limb SUMER images do not contain any features (e.g., limb brightening), the coaligned EIS and SUMER images might have a residual offset up to $10\arcsec$ to $15\arcsec$ in the north\textendash south direction.

\section{Spicules, stray light, and Opacity Effects}\label{appen:stray_los}

\begin{figure}
    \centering
    \includegraphics[width=0.6\linewidth]{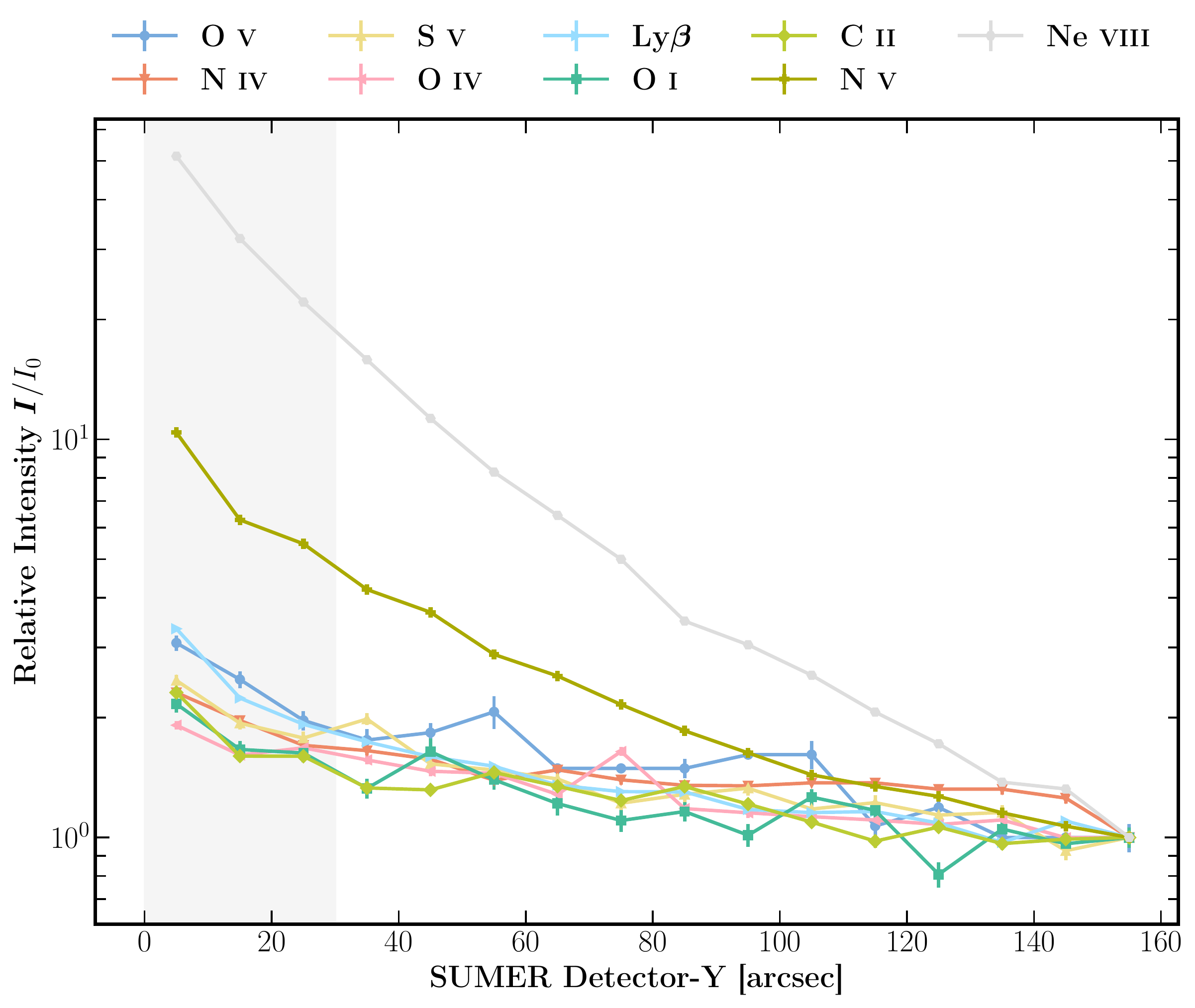}
    \caption{Normalized intensity of the cooler stray light lines (O \textsc{i}, Ly$\beta$, C \textsc{ii}, O \textsc{iv}, O \textsc{v}, N \textsc{iv}, and S \textsc{v}), and hotter lines (N \textsc{v} and Ne \textsc{viii}) along the SUMER slit. The vertical shaded area represents the averaged 30 pixels. Link to the \texttt{Jupyter} notebook creating this figure \href{https://github.com/yjzhu-solar/EIS_SUMER_PCH_Ti/blob/main/ipynb/sumer_fit/sumer_stray_intensity.ipynb}{\faGithub}.}
    \label{fig:sumer_stray_int}
\end{figure}

In this study, we assumed that the spectral lines observed by SUMER and EIS are optically thin and are emitted by the plasma in the coronal hole. However, as we analyzed the observation between 1.01 and 1.04\,$R_\odot$, spicules might have contributed to the emission. To address the potential cold plasma along the LOS, we show the intensity distribution of the chromospheric and TR lines along the SUMER slit in Figure~\ref{fig:sumer_stray_int}. We found that the intensity of these cooler lines gradually decreases with height, like the common stray light found in other SUMER off-limb observations. The observed stray-light lines do not show significant enhancement at the bottom of the slit. And at the height we used for this work (pixels 0 to 30), the stray light is negligible. On the other hand, the intensities of hotter lines like N \textsc{v} and Ne \textsc{viii} decrease drastically along the slit. Therefore, we conclude that the cold chromospheric and TR plasma, such as spicules, does not significantly contaminate the line profiles used for $T_i$ diagnostics.    

\citet{DelZanna2019} found the anomalous variation of the Fe \textsc{xii} 195/192 ratio in the off-limb quiet-Sun corona. The Fe \textsc{xii} 195/192 ratio should not be sensitive to density or temperature. They suggested the reduction in Fe \textsc{xii} 195/192 ratio is caused by the optical thickness in the strongest Fe \textsc{xii} 195\,\mbox{\AA}. We measured $I_{195}/I_{192} = 3.25$ between 1.01 and 1.04 $R_\odot$ using the GDZ calibration, which is fairly close to the optically thin limit $I_{195}/I_{192} = 3.15$ given by CHIANTI. Hence, we suggested that the optical thickness in Fe \textsc{xii} 195\,\mbox{\AA} is still negligible in our observations.

\section{Wavelength Dependence of the EIS instrumental Widths}\label{append:dlamb_EIS_lamb} 
The latest EIS instrumental width provided by the EIS software in SolarSoft is constant at different wavelengths. However, earlier studies of the EIS instrumental width suggested that the instrumental widths in the two detectors are slightly different \citep[e.g.,][]{Brown2008}. We used the following method to investigate whether the EIS instrumental width depends on the wavelength. The fitted FHWM $\Delta \lambda_{\rm fit}$ is often interpreted as 
\begin{equation}
    \Delta \lambda_{\rm fit}^2 = \Delta \lambda_{\rm inst}^2 + 4\ln 2 \frac{v_{\rm eff}^2}{c^2} \lambda_0^2  
\end{equation}
where $\Delta \lambda_{\rm inst}$ is the instrumental FWHM. Assuming that the effective velocity $v_{\rm eff}^2 = 2k_B T_i/m_i + \xi^2 $ is a constant for all spectral lines from the same ion, we can treat $\Delta \lambda_{\rm fit}$ as a function of $\lambda_0$ with two parameters $\Delta \lambda_{\rm inst}$ and $v_{\rm eff}$, i.e., $\Delta \lambda_{\rm fit} = f\left(\lambda_0;\Delta \lambda_{\rm inst},v_{\rm eff}\right)$. If $\Delta \lambda_{\rm inst}$ does not depend on the wavelength, we could use $(\Delta \lambda_{\rm fit},\lambda_0)$ pairs from different spectral lines of the same ion to fit the optimized $\Delta \lambda_{\rm inst}$ and $v_{\rm eff}$. 

\begin{figure*}[htb!]
    \centering
    \includegraphics[width=0.8\linewidth]{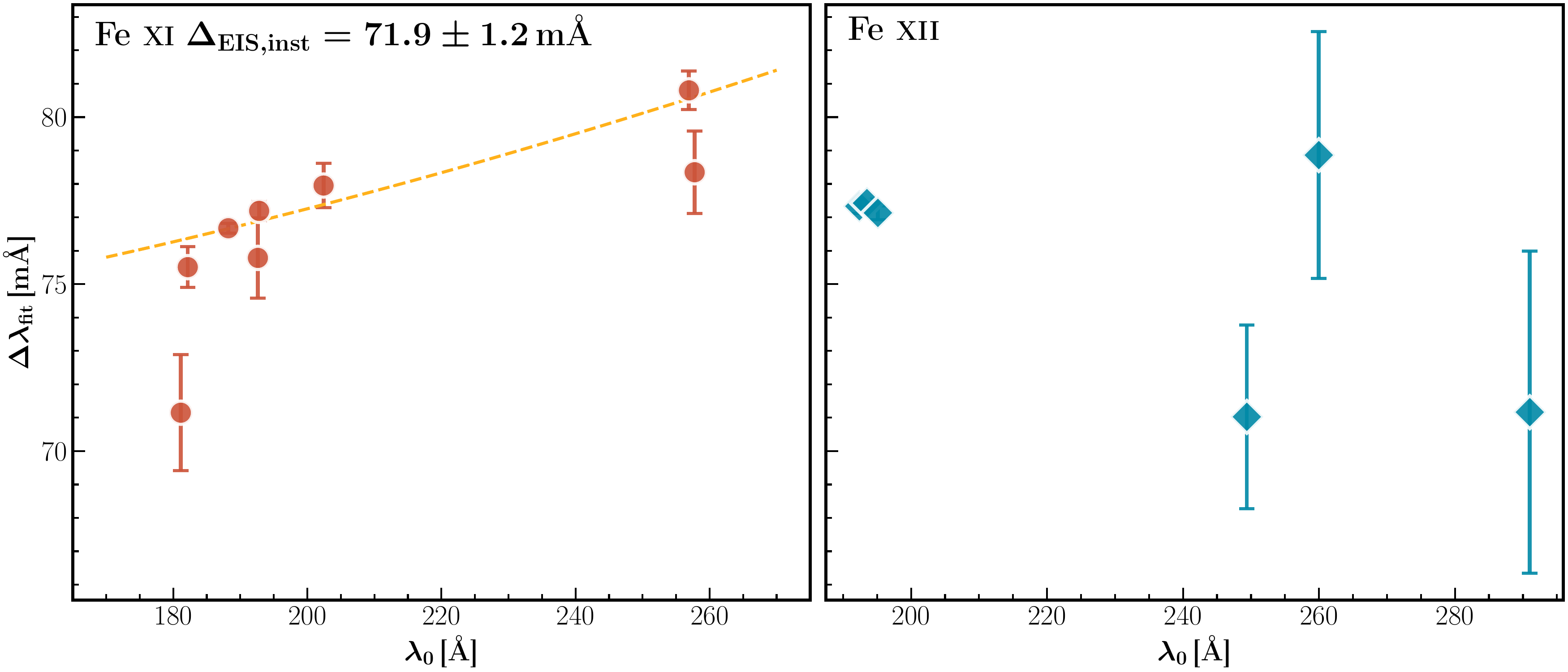}
    \caption{Fitted FWHM $\Delta \lambda_{\rm fit}$ versus line centroid wavelength $\lambda_0$ of the Fe \textsc{xi} and Fe \textsc{xii} lines in off-limb quiet-Sun corona. The dashed line in the left panel shows the best fitting of $\Delta \lambda_{\rm fit} = f\left(\lambda_0|\Delta \lambda_{\rm inst},v_{\rm eff}\right)$. Link to the \texttt{Jupyter} notebook creating this figure \href{https://github.com/yjzhu-solar/EIS_SUMER_PCH_Ti/blob/main/ipynb/paper/eis_dlamb_inst_lamb.ipynb}{\faGithub}.}
    \label{fig:dlamb_lamb}
\end{figure*}

We implemented this method on an EIS observation of the west off-limb quiet-Sun corona on 2007 April 13. The data set has been studied in \citet{Landi2010} to cross-calibrate the intensity between EIS and SUMER, and we chose it because of its high S/N. We averaged the data at the same 30 pixels on the CCD detector used in this study. Although there are barely isolated and strong lines of the same ion across the EIS detector, we found the Fe \textsc{xi} and Fe \textsc{xii} lines are the best candidates. Figure~\ref{fig:dlamb_lamb} shows the fitted FWHM $\Delta \lambda_{\rm fit}$ and the line centroid wavelength $\lambda_0$ of the Fe \textsc{xi} and Fe \textsc{xii} lines. For Fe \textsc{xi} lines, we obtained an instrumental width $\Delta \lambda_{\rm inst} = 71.9\pm1.2$\,m\mbox{\AA}, which is more consistent with the instrumental width $\Delta \lambda_{\rm inst} = 69.7$\,m\mbox{\AA} given by the EIS software. However, there are some outliers in the Fe \textsc{xi} lines, including Fe \textsc{xi} 181.130\,\mbox{\AA} and 257.772\,\mbox{\AA}. The Fe \textsc{xii} triplets at 192.394, 193.509, and 195.119,\mbox{\AA} have very similar line widths $\Delta \lambda_{\rm fit} \sim 77$\,m\mbox{\AA}. However, the $\Delta \lambda_{\rm fit}$ of the Fe \textsc{xii} 249.388, 259.973, and 291.010\,\mbox{\AA} line does show a monotonic dependence on $\lambda_0$. Therefore, we cannot obtain the instrumental width from the Fe \textsc{xii} widths. We suggest that the instrumental width of the EIS $2\arcsec$ slit might depend on the wavelength. 

\section{Line-Fitting Examples}\label{appen:fit_example}
\begin{figure*}
\gridline{\fig{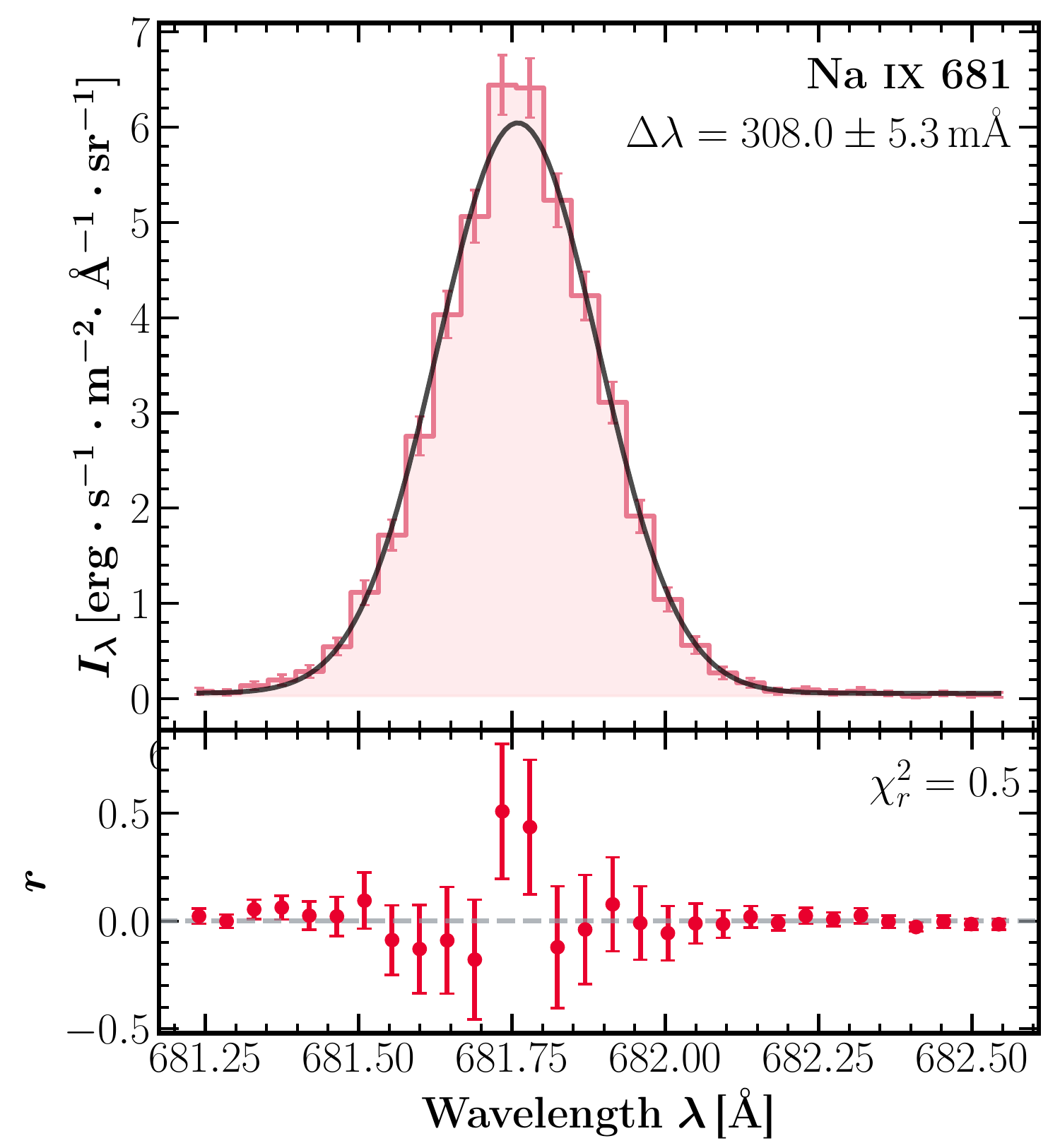}{0.40\textwidth}{(a)}
          \fig{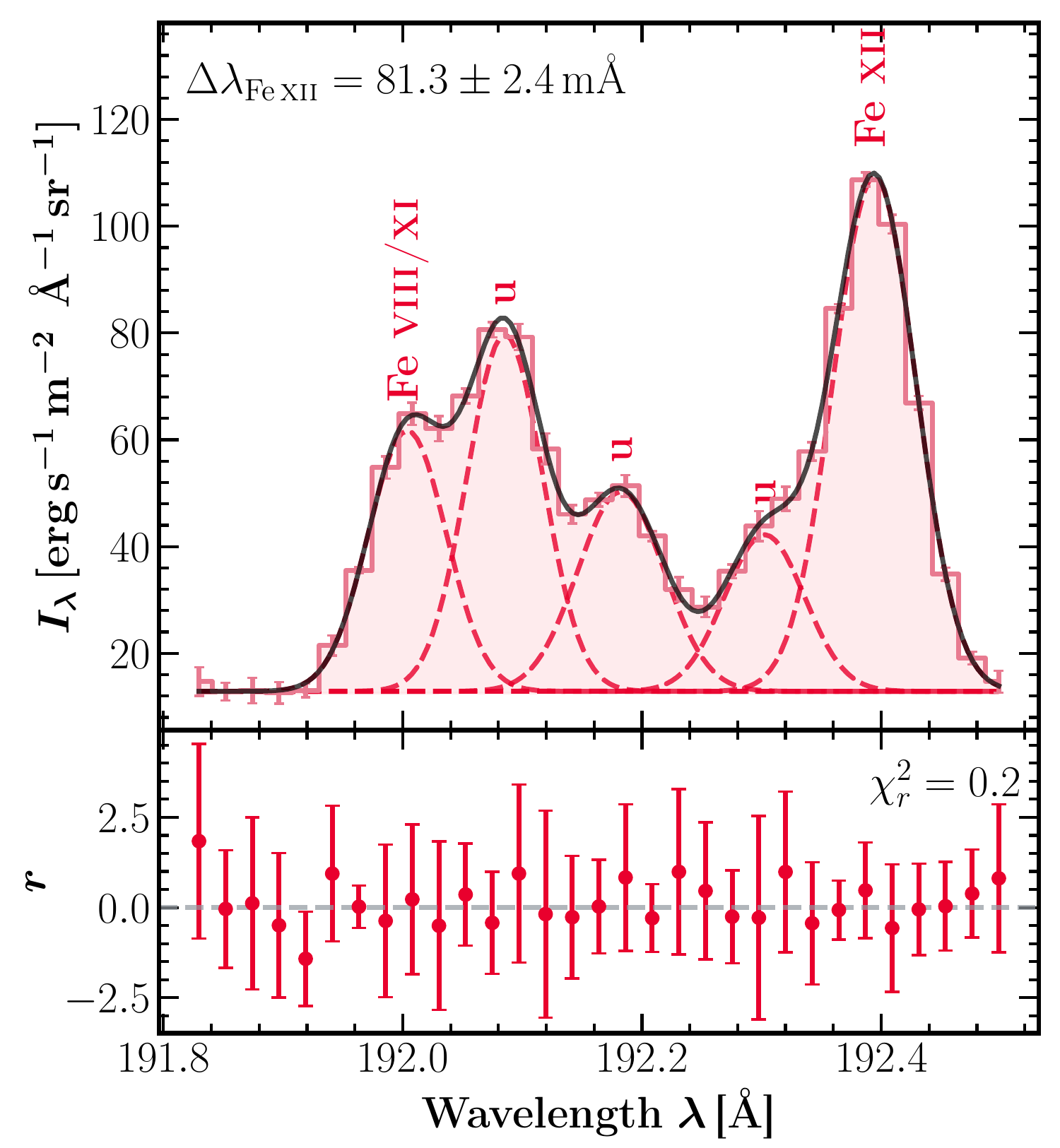}{0.40\textwidth}{(b)}}
    \caption{Examples of the single and multi-Gaussian fitting. (a) Single-Gaussian fitting of the Na \textsc{ix} 681\,\mbox{\AA} line observed by SUMER. (b) Multi-Gaussian fitting of the blended lines near 192\,\mbox{\AA} observed by EIS, including the blended Fe \textsc{viii} and Fe \textsc{xi} 192.021\,\mbox{\AA}, Fe \textsc{xii} 192.394\,\mbox{\AA}, and three other unidentified (u) lines. The top panel shows the observed spectrum (step line) and the fitting line profile (solid black line). The lower panel displays the fitting residual. Links to the \texttt{Jupyter} notebook creating panels (a) \href{https://github.com/yjzhu-solar/EIS_SUMER_PCH_Ti/blob/main/ipynb/sumer_fit/window_338_NaIX_example.ipynb}{\faGithub} and (b) \href{https://github.com/yjzhu-solar/EIS_SUMER_PCH_Ti/blob/main/ipynb/eis_fit/eis_fit_192_example.ipynb}{\faGithub}.}
    \label{fig:fit_example}
\end{figure*}

Figure~\ref{fig:fit_example}(a) displays the single-Gaussian fitting of the Na~\textsc{ix} 681\,\mbox{\AA} line observed by SUMER. The line wings resolved by SUMER are well fitted by the Gaussian function. On the other hand, the peak intensity of the observed line core is larger than the best-fit line profile. A multi-Gaussian fitting example of several spectral lines around 192\,\mbox{\AA} observed by EIS is shown in Figure~\ref{fig:fit_example}(b). Although the Fe \textsc{xii} 192.394\,\mbox{\AA} line is blended with some unidentified lines in the blue wing, we can still get a proper fitting of line widths. Note that the reduced $\chi^2$ function in both examples is less than 1, which might indicate that the Poisson statistics or propagation of errors overestimates the uncertainty. 

\section{Table of Line Widths}
\begin{longrotatetable}
\begin{deluxetable*}{@{\extracolsep{2pt}}ccccccccccccc}
\def\arraystretch{1.4}
\tabletypesize{\scriptsize}
\tablecolumns{13}
\tablewidth{0pt}
\tablecaption{Estimated ranges of ion temperatures $T_{i,\mathrm{min}}$ and $T_{i,\mathrm{max}}$, average effective velocity $\bar{v}_{\rm eff}$, fitted line width $\Delta \lambda_{\rm fit}$, and corrected width $\Delta \lambda_{\rm true}$.}

\tablehead
{
\colhead{} & \colhead{} & \colhead{} & \multicolumn{6}{c}{Instrument Averaged Values} & \multicolumn{4}{c}{Individual Line Measurements} \\\cline{4-9} \cline{10-13}
\colhead{Ion} & \colhead{$Z/A$} & \colhead{Inst.} &  \colhead{$\log T_{i,\mathrm{min}}$\,[K]} & \colhead{$\log T_{i,\mathrm{max}}$\,[K]} & \colhead{$\log T^*_{i,\mathrm{min}}$\,[K]} & \colhead{$\log T^*_{i,\mathrm{max}}$\,[K]} & \colhead{$\bar{v}_{\rm eff}\,\mathrm{\left[km\, s^{-1} \right]}$} & \colhead{$\bar{v}^*_{\rm eff}\,\mathrm{\left[km\, s^{-1} \right]}$} & \colhead{$\lambda$\,[\mbox{\AA}]} & \colhead{$\Delta \lambda_{\rm true}$\,[m\mbox{\AA}]} & \colhead{$\Delta \lambda^*_{\rm true}$\,[m\mbox{\AA}]} & \colhead{$\Delta \lambda_{\rm fit}$\,[m\mbox{\AA}]}
} 
\startdata
\multirow{5}{*}{Fe \textsc{viii}} & \multirow{5}{*}{0.125} & \multirow{4}{*}{EIS}   & \multirow{4}{*}{\dots\dots\dots}        & \multirow{4}{*}{$6.54^{+0.04}_{-0.05}$} & \multirow{4}{*}{$6.44^{+0.06}_{-0.08}$} & \multirow{4}{*}{$6.79^{+0.03}_{-0.03}$} & \multirow{4}{*}{$32.0\pm 1.7$}                             & \multirow{4}{*}{$43.0\pm 1.5$}                             & 185.213                 & $31.4\pm1.1$                               & $44.1\pm0.8$                               & $76.48\pm0.45$                            \\
                                  &                        &                        &                                         &                                         &                                         &                                         &                                                            &                                                            & 186.598                 & $38.8\pm1.5$                               & $49.7\pm1.1$                               & $79.79\pm0.71$                            \\
                                  &                        &                        &                                         &                                         &                                         &                                         &                                                            &                                                            & 194.661                 & $33.1\pm1.1$                               & $45.3\pm0.8$                               & $77.16\pm0.49$                            \\
                                  &                        &                        &                                         &                                         &                                         &                                         &                                                            &                                                            & 253.956                 & $39.6\pm5.6$                               & $49.7\pm 4.5$                              & $79.8\pm2.8$                              \\ \cline{3-13} 
                                  &                        & SUMER                  & $6.70^{+0.18}_{-0.30}$                  & $6.93^{+0.11}_{-0.15}$                  & \dots\dots\dots                         & \dots\dots\dots                         & $50.2\pm 7.5$                                              & \dots\dots\dots                                            & 697.156                 & $194\pm29$                                 & \dots\dots\dots                            & $281\pm20$                                \\ \hline
Mn \textsc{viii}                  & 0.127                  & EIS                    & $6.45^{+0.23}_{-0.53}$                  & $6.79^{+0.12}_{-0.17}$                  & $6.75^{+0.13}_{-0.19}$                  & $6.96^{+0.09}_{-0.11}$                  & $43.4\pm 7.0$                                              & $52.5\pm5.8$                                               & 185.455                 & $44.7\pm7.2$                               & $54.4\pm 5.9$                              & $82.8\pm3.9$                              \\ \hline
\multirow{2}{*}{Fe \textsc{ix}}   & \multirow{2}{*}{0.143} & \multirow{2}{*}{EIS}   & \multirow{2}{*}{$6.08^{+0.08}_{-0.10}$} & \multirow{2}{*}{$6.67^{+0.02}_{-0.02}$} & \multirow{2}{*}{$6.57^{+0.02}_{-0.03}$} & \multirow{2}{*}{$6.86^{+0.01}_{-0.01}$} & \multirow{2}{*}{$37.2\pm 1.0$}                             & \multirow{2}{*}{$46.3\pm 0.7$}                             & 191.206                 & $33.1\pm4.2$                               & $45.2\pm3.1$                               & $77.1\pm1.8$                              \\
                                  &                        &                        &                                         &                                         &                                         &                                         &                                                            &                                                            & 197.854                 & $41.0\pm0.7$                               & $51.3\pm0.6$                               & $80.85\pm0.35$                            \\ \hline
\multirow{7}{*}{Fe \textsc{x}}    & \multirow{7}{*}{0.161} & \multirow{5}{*}{EIS}   & \multirow{5}{*}{$6.06^{+0.08}_{-0.09}$} & \multirow{5}{*}{$6.66^{+0.02}_{-0.02}$} & \multirow{5}{*}{$6.61^{+0.03}_{-0.03}$} & \multirow{5}{*}{$6.88^{+0.02}_{-0.02}$} & \multirow{5}{*}{$37.0\pm 0.9$}                             & \multirow{5}{*}{$47.4\pm 0.9$}                             & 174.531                 & $38.0\pm3.1$                               & $49.0\pm2.4$                               & $79.4\pm1.5$                              \\
                                  &                        &                        &                                         &                                         &                                         &                                         &                                                            &                                                            & 177.240                 & $40.7\pm1.3$                               & $51.2\pm1.0$                               & $80.8\pm0.6$                              \\
                                  &                        &                        &                                         &                                         &                                         &                                         &                                                            &                                                            & 184.537                 & $37.1\pm0.6$                               & $48.4\pm0.4$                               & $79.0\pm0.3$                              \\
                                  &                        &                        &                                         &                                         &                                         &                                         &                                                            &                                                            & 190.037                 & $38.0\pm1.0$                               & $50.4\pm0.8$                               & $80.3\pm0.5$                              \\
                                  &                        &                        &                                         &                                         &                                         &                                         &                                                            &                                                            & 193.715                 & $34.6\pm2.7$                               & $46.4\pm2.0$                               & $77.8\pm1.2$                              \\ \cline{3-13} 
                                  &                        & \multirow{2}{*}{SUMER} & \multirow{2}{*}{$6.77^{+0.06}_{-0.06}$} & \multirow{2}{*}{$6.97^{+0.04}_{-0.04}$} & \multirow{2}{*}{\dots\dots\dots}        & \multirow{2}{*}{\dots\dots\dots}        & \multirow{2}{*}{$52.7\pm 2.3$}                             & \multirow{2}{*}{\dots\dots\dots}                           & 1028.053                & \multirow{2}{*}{$301\pm13$}                & \multirow{2}{*}{\dots\dots\dots}           & \multirow{2}{*}{$355\pm11$}               \\
                                  &                        &                        &                                         &                                         &                                         &                                         &                                                            &                                                            & 1028.082                &                                            &                                            &                                           \\ \hline
Ar \textsc{viii}                  & 0.175                  & SUMER                  & $6.54^{+0.13}_{-0.19}$                  & $6.77^{+0.08}_{-0.10}$                  & \dots\dots\dots                         & \dots\dots\dots                         & $49.6\pm5.1$                                               & \dots\dots\dots                                            & 713.801                 & $197\pm20$                                 & \dots\dots\dots                            & $282\pm14$                                \\ \hline
\multirow{6}{*}{Fe \textsc{xi}}   & \multirow{6}{*}{0.179} & \multirow{5}{*}{EIS}   & \multirow{5}{*}{$6.17^{+0.03}_{-0.03}$} & \multirow{5}{*}{$6.69^{+0.01}_{-0.01}$} & \multirow{5}{*}{$6.63^{+0.01}_{-0.01}$} & \multirow{5}{*}{$6.89^{+0.01}_{-0.01}$} & \multirow{5}{*}{$38.3 \pm 0.4$}                            & \multirow{5}{*}{$48.1\pm 0.3$}                             & 182.167                 & $39.6\pm3.0$                               & $50.3\pm2.4$                               & $80.2\pm1.5$                              \\
                                  &                        &                        &                                         &                                         &                                         &                                         &                                                            &                                                            & 188.216                 & $39.9\pm0.7$                               & $50.5\pm0.5$                               & $80\pm0.3$                                \\
                                  &                        &                        &                                         &                                         &                                         &                                         &                                                            &                                                            & 188.299                 & $39.9\pm0.7$                               & $50.5\pm0.5$                               & $80\pm0.3$                                \\
                                  &                        &                        &                                         &                                         &                                         &                                         &                                                            &                                                            & 192.627                 & $42.7\pm3.1$                               & $52.7\pm2.5$                               & $81.7\pm1.6$                              \\
                                  &                        &                        &                                         &                                         &                                         &                                         &                                                            &                                                            & 202.424                 & $51.4\pm5.1$                               & $60.0\pm4.3$                               & $86.6\pm3.0$                              \\ \cline{3-13} 
                                  &                        & SUMER                  & $6.55^{+0.19}_{-0.36}$                  & $6.84^{+0.11}_{-0.15}$                  & \dots\dots\dots                         & \dots\dots\dots                         & $45.6\pm 6.5$                                              & \dots\dots\dots                                            & 1028.955                & $260\pm37$                                 & \dots\dots\dots                            & $322\pm30$                                \\ \hline
\multirow{3}{*}{Fe \textsc{xii}}  & \multirow{3}{*}{0.197} & \multirow{3}{*}{EIS}   & \multirow{3}{*}{$6.35^{+0.10}_{-0.14}$} & \multirow{3}{*}{$6.75^{+0.04}_{-0.05}$} & \multirow{3}{*}{$6.69^{+0.05}_{-0.06}$} & \multirow{3}{*}{$6.92^{+0.03}_{-0.03}$} & \multirow{3}{*}{$41.1\pm 2.2$}                             & \multirow{3}{*}{$49.8\pm 1.8$}                             & 192.394                 & $41.9\pm4.7$                               & $52.1\pm3.7$                               & $81.3\pm2.4$                              \\
                                  &                        &                        &                                         &                                         &                                         &                                         &                                                            &                                                            & 193.509                 & $40.4\pm1.0$                               & $50.9\pm0.8$                               & $80.6\pm0.5$                              \\
                                  &                        &                        &                                         &                                         &                                         &                                         &                                                            &                                                            & 195.119                 & $47.5\pm0.9$                               & $56.7\pm0.8$                               & $84.4\pm0.5$                              \\ \hline
Mg \textsc{vi}                    & 0.206                  & EIS                    & $5.63^{+0.62}_{-}$                      & $6.28^{+0.23}_{-0.55}$                  & $5.99^{+0.38}_{-}$                      & $6.40^{+0.19}_{-0.35}$                  & $36.3\pm 13.0$                                             & $41.3\pm 11.4$                                             & 268.991                 & $54\pm19$                                  & $62.0\pm17.0$                              & $88\pm12$                                 \\ \hline
\multirow{4}{*}{Si \textsc{vii}}  & \multirow{4}{*}{0.214} & \multirow{3}{*}{EIS}   & \multirow{3}{*}{$5.94^{+0.02}_{-0.02}$} & \multirow{3}{*}{$6.41^{+0.01}_{-0.01}$} & \multirow{3}{*}{$6.42^{+0.10}_{-0.13}$} & \multirow{3}{*}{$6.64^{+0.06}_{-0.07}$} & \multirow{3}{*}{$39.2\pm 0.3$}                             & \multirow{3}{*}{$43.7\pm 0.2$}                             & 272.647                 & $59.4\pm2.3$                               & $66.5\pm2.1$                               & $91.2\pm1.5$                              \\
                                  &                        &                        &                                         &                                         &                                         &                                         &                                                            &                                                            & 275.361                 & $60.1\pm0.8$                               & $67.1\pm0.7$                               & $91.7\pm0.5$                              \\
                                  &                        &                        &                                         &                                         &                                         &                                         &                                                            &                                                            & 275.675                 & $56.9\pm4.1$                               & $64.3\pm3.6$                               & $89.6\pm2.6$                              \\ \cline{3-13} 
                                  &                        & SUMER                  & $6.42^{+0.10}_{-0.13}$                  & $6.64^{+0.06}_{-0.07}$                  & \dots\dots\dots                         & \dots\dots\dots                         & $50.8\pm 3.9$                                              & \dots\dots\dots                                            & 1049.153                & $296\pm23$                                 & \dots\dots\dots                            & $352\pm19$                                \\ \hline
Fe \textsc{xiii}                  & 0.215                  & EIS                    & $6.21^{+0.13}_{-0.19}$                  & $6.70^{+0.05}_{-0.05}$                  & $6.61^{+0.06}_{-0.07}$                  & $6.88^{+0.03}_{-0.03}$                  & $38.8\pm 2.2$                                              & $47.4\pm 1.8$                                              & 202.044                 & $43.5\pm2.5$                               & $53.4\pm2.0$                               & $82.2\pm1.3$                              \\ \hline
Al \textsc{vii}                   & 0.222                  & SUMER                  & $6.06^{+0.26}_{-0.77}$                  & $6.45^{+0.13}_{-0.18}$                  & \dots\dots\dots                         & \dots\dots\dots                         & $41.7\pm 7.1$                                              & \dots\dots\dots                                            & 1053.996                & $244\pm42$                                 & \dots\dots\dots                            & $307\pm33$                                \\ \hline
Mg \textsc{vii}                   & 0.247                  & EIS                    & $5.84^{+0.18}_{-0.32}$                  & $6.34^{+0.07}_{-0.08}$                  & $6.08^{+0.12}_{-0.16}$                  & $6.43^{+0.05}_{-0.06}$                  & $38.7\pm3.2$                                               & $43.1\pm 2.9$                                              & 276.154                 & $59.3\pm4.9$                               & $66.3\pm4.4$                               & $91.1\pm3.2$                              \\ \hline
\multirow{3}{*}{S \textsc{x}}     & \multirow{3}{*}{0.281} & \multirow{2}{*}{EIS}   & \multirow{2}{*}{$5.85^{+0.20}_{-0.39}$} & \multirow{2}{*}{$6.43^{+0.06}_{-0.07}$} & \multirow{2}{*}{$6.16^{+0.12}_{-0.16}$} & \multirow{2}{*}{$6.54^{+0.05}_{-0.06}$} & \multirow{2}{*}{$37.3\pm 2.9$}                             & \multirow{2}{*}{$42.3\pm 2.7$}                             & 259.496                 & $61\pm15$                                  & $67.6\pm14$                                & $92\pm10$                                 \\
                                  &                        &                        &                                         &                                         &                                         &                                         &                                                            &                                                            & 264.230                 & $52.0\pm9.7$                               & $60.0\pm8.4$                               & $86.6\pm5.8$                              \\ \cline{3-13} 
                                  &                        & SUMER                  & $6.86^{+0.19}_{-0.35}$                  & $6.97^{+0.16}_{-0.25}$                  & \dots\dots\dots                         & \dots\dots\dots                         & $69.4\pm 15.0$                                             & \dots\dots\dots                                            & 776.373                 & $299\pm65$                                 & \dots\dots\dots                            & $358\pm54$                                \\ \hline
\multirow{2}{*}{Si \textsc{ix}}   & \multirow{2}{*}{0.285} & EIS                    & $5.55^{+0.92}_{-}$                      & $6.32^{+0.35}_{-}$                      & $6.04^{+0.52}_{-}$                      & $6.46^{+0.28}_{-0.96}$                  & $35.1\pm 21.5$                                             & $41.1\pm 18.3$                                             & 258.080                 & $50\pm31$                                  & $59.1\pm26.2$                              & $86\pm18$                                 \\ \cline{3-13} 
                                  &                        & SUMER                  & $5.75^{+0.36}_{-}$                      & $6.36^{+0.12}_{-0.17}$                  & \dots\dots\dots                         & \dots\dots\dots                         & $36.8\pm 5.9$                                              & \dots\dots\dots                                            & 694.686                 & $142\pm23$                                 & \dots\dots\dots                            & $249\pm13$                                \\ \hline
\multirow{4}{*}{Mg \textsc{viii}} & \multirow{4}{*}{0.288} & \multirow{4}{*}{SUMER} & \multirow{4}{*}{$6.32^{+0.05}_{-0.05}$} & \multirow{4}{*}{$6.55^{+0.03}_{-0.03}$} & \multirow{4}{*}{\dots\dots\dots}        & \multirow{4}{*}{\dots\dots\dots}        & \multirow{4}{*}{$49.4\pm 1.7$}                             & \multirow{4}{*}{\dots\dots\dots}                           & 689.641                 & $187\pm27$                                 & \dots\dots\dots                            & $276\pm18$                                \\
                                  &                        &                        &                                         &                                         &                                         &                                         &                                                            &                                                            & 762.660                 & $163\pm38$                                 & \dots\dots\dots                            & $258\pm24$                                \\
                                  &                        &                        &                                         &                                         &                                         &                                         &                                                            &                                                            & 772.260                 & $197\pm12$                                 & \dots\dots\dots                            & $280.2\pm8.6$                             \\
                                  &                        &                        &                                         &                                         &                                         &                                         &                                                            &                                                            & 782.362                 & $222.9\pm7.8$                              & \dots\dots\dots                            & $298.4\pm5.8$                             \\ \hline
Al \textsc{ix}                    & 0.296                  & SUMER                  & $5.91^{+0.83}_{-}$                      & $6.39^{+0.46}_{-}$                      & \dots\dots\dots                         & \dots\dots\dots                         & $39\pm 37$                                                 & \dots\dots\dots                                            & 703.73                  & $154\pm143$                                & \dots\dots\dots                            & $256\pm86$                                \\ \hline
\multirow{3}{*}{O \textsc{vi}}    & \multirow{3}{*}{0.313} & EIS                    & $5.78^{+0.11}_{-0.15}$                  & $6.20^{+0.05}_{-0.05}$                  & $6.16^{+0.05}_{-0.06}$                  & $6.39^{+0.03}_{-0.03}$                  & $40.6\pm 2.3$                                              & $50.3\pm 1.9$                                              & 184.117                 & $41.5\pm2.3$                               & $51.7\pm1.9$                               & $81.1\pm1.2$                              \\ \cline{3-13} 
                                  &                        & \multirow{2}{*}{SUMER} & \multirow{2}{*}{$6.16^{+0.01}_{-0.01}$} & \multirow{2}{*}{$6.39^{+0.01}_{-0.01}$} & \multirow{2}{*}{\dots\dots\dots}        & \multirow{2}{*}{\dots\dots\dots}        & \multirow{2}{*}{$50.3\pm 0.4$}                             & \multirow{2}{*}{\dots\dots\dots}                           & 1031.912                & $294.4\pm3.3$                              & \dots\dots\dots                            & $350.2\pm2.8$                             \\
                                  &                        &                        &                                         &                                         &                                         &                                         &                                                            &                                                            & 1037.613                & $288.8\pm1.2$                              & \dots\dots\dots                            & $345.6\pm1.0$                             \\ \hline
\multirow{3}{*}{Si \textsc{x}}    & \multirow{3}{*}{0.320} & \multirow{2}{*}{EIS}   & \multirow{2}{*}{$5.76^{+0.12}_{-0.16}$} & \multirow{2}{*}{$6.36^{+0.03}_{-0.03}$} & \multirow{2}{*}{$6.09^{+0.06}_{-0.07}$} & \multirow{2}{*}{$6.48^{+0.03}_{-0.03}$} & \multirow{2}{*}{$36.9\pm 1.4$}                             & \multirow{2}{*}{$42.1\pm 1.3$}                             & 258.374                 & $54.0\pm4.2$                               & $61.7\pm3.7$                               & $87.8\pm2.6$                              \\
                                  &                        &                        &                                         &                                         &                                         &                                         &                                                            &                                                            & 261.056                 & $49.5\pm8.1$                               & $57.8\pm6.9$                               & $85.1\pm4.7$                              \\ \cline{3-13} 
                                  &                        & SUMER                  & \dots\dots\dots                         & $6.24^{+0.17}_{-0.27}$                  & \dots\dots\dots                         & \dots\dots\dots                         & $32.2\pm 7.5$                                              & \dots\dots\dots                                            & 624.694                 & $112\pm26$                                 & \dots\dots\dots                            & $145\pm20$                                \\ \hline
\multirow{2}{*}{Mg \textsc{ix}}   & \multirow{2}{*}{0.329} & \multirow{2}{*}{SUMER} & \multirow{2}{*}{$6.53^{+0.03}_{-0.04}$} & \multirow{2}{*}{$6.69^{+0.02}_{-0.02}$} & \multirow{2}{*}{\dots\dots\dots}        & \multirow{2}{*}{\dots\dots\dots}        & \multirow{2}{*}{$57.6\pm 1.6$}                             & \multirow{2}{*}{\dots\dots\dots}                           & 706.060                 & $228.5\pm 4.7$                             & \dots\dots\dots                            & $305.0\pm3.5$                             \\
                                  &                        &                        &                                         &                                         &                                         &                                         &                                                            &                                                            & 749.552                 & $223\pm13$                                 & \dots\dots\dots                            & $300.9\pm9.2$                             \\ \hline
\multirow{2}{*}{Ne \textsc{viii}} & \multirow{2}{*}{0.347} & \multirow{2}{*}{SUMER} & \multirow{2}{*}{$6.36^{+0.02}_{-0.02}$} & \multirow{2}{*}{$6.55^{+0.01}_{-0.01}$} & \multirow{2}{*}{\dots\dots\dots}        & \multirow{2}{*}{\dots\dots\dots}        & \multirow{2}{*}{$53.8\pm 0.9$}                             & \multirow{2}{*}{\dots\dots\dots}                           & 770.428                 & $228.2\pm2.3$                              & \dots\dots\dots                            & $302.3\pm1.7$                             \\
                                  &                        &                        &                                         &                                         &                                         &                                         &                                                            &                                                            & 780.385                 & $239.8\pm3.9$                              & \dots\dots\dots                            & $311.0\pm3.0$                             \\ \hline
Na \textsc{ix}                    & 0.348                  & SUMER                  & $6.58^{+0.04}_{-0.04}$                  & $6.72^{+0.03}_{-0.03}$                  & \dots\dots\dots                         & \dots\dots\dots                         & $61.4\pm 1.9$                                              & \dots\dots\dots                                            & 681.719                 & $232.5\pm7.2$                              & \dots\dots\dots                            & $308.0\pm5.4$                             \\ \hline
\multirow{2}{*}{Mg \textsc{x}}    & \multirow{2}{*}{0.370} & \multirow{2}{*}{SUMER} & \multirow{2}{*}{$6.17^{+0.11}_{-0.14}$} & \multirow{2}{*}{$6.48^{+0.06}_{-0.06}$} & \multirow{2}{*}{\dots\dots\dots}        & \multirow{2}{*}{\dots\dots\dots}        & \multirow{2}{*}{$45.2\pm 3.1$}                             & \multirow{2}{*}{\dots\dots\dots}                           & 609.793                 & $145.1\pm2.6$                              & \dots\dots\dots                            & $171.4\pm2.2$                             \\
                                  &                        &                        &                                         &                                         &                                         &                                         &                                                            &                                                            & 624.941                 & $170.6\pm3.4$                              & \dots\dots\dots                            & $193.3\pm3.0$                             \\ \hline
\enddata

\tablecomments{Quantities with $^*$ are calculated using the EIS instrumental width $\Delta \lambda'_{\rm inst,EIS} = 62.7\,$m\mbox{\AA} cross-calibrated from SUMER observations. The fitting of the individual spectral line can be found in the following \texttt{Jupyter} notebooks: \href{https://github.com/yjzhu-solar/EIS_SUMER_PCH_Ti/blob/main/ipynb/eis_fit/eis_fit.ipynb}{EIS}, \href{https://github.com/yjzhu-solar/EIS_SUMER_PCH_Ti/blob/main/ipynb/sumer_fit/window_338_fit_poisson.ipynb}{SUMER window 1}, \href{https://github.com/yjzhu-solar/EIS_SUMER_PCH_Ti/blob/main/ipynb/sumer_fit/window_375_fit_poisson.ipynb}{SUMER window 2}, \href{https://github.com/yjzhu-solar/EIS_SUMER_PCH_Ti/blob/main/ipynb/sumer_fit/window_509_fit_poisson.ipynb}{SUMER window 3}, and \href{https://github.com/yjzhu-solar/EIS_SUMER_PCH_Ti/blob/main/ipynb/sumer_fit/window_607_fit_poisson.ipynb}{SUMER window 4}. \label{tab:line_width}}

\end{deluxetable*}
\end{longrotatetable}

\bibliography{ms}{}
\bibliographystyle{aasjournal}


\end{document}